# Characterizing the reproduction number of epidemics with early sub-exponential growth dynamics

Gerardo Chowell[1,2], Cécile Viboud[2], Lone Simonsen[3,4], Seyed M. Moghadas[5]

[1] School of Public Health, Georgia State University, Atlanta, GA, USA

[2] Division of International Epidemiology and Population Studies, Fogarty International Center, National Institutes of Health, Bethesda, MD, USA

[3] Department of Public health, University of Copenhagen, Copenhagen, Denmark

[4] Department of Global Health, George Washington University, Washington DC, USA

[5] Agent Based Modelling Laboratory, York University, Toronto, Canada.

**Corresponding author:**

Gerardo Chowell, PhD
School of Public Health, Georgia State University, Atlanta, GA, USA
Division of International Epidemiology and Population Studies, Fogarty International Center, National Institutes of Health, Bethesda, MD, USA
Email: gchowell@gsu.edu

Body word count: 4,531

Abstract word count: 314

**Abstract**

Early estimates of the transmission potential of emerging and re-emerging infections are increasingly used to inform public health authorities on the level of risk posed by outbreaks. Existing methods to estimate the reproduction number generally assume exponential growth in case incidence in the first few disease generations, before susceptible depletion sets in. In reality, outbreaks can display sub-exponential (i.e., polynomial) growth in the first few disease generations, owing to clustering in contact patterns, spatial effects, inhomogeneous mixing, reactive behavior changes, or other mechanisms. Here, we introduce the *generalized* growth model to characterize the early growth profile of outbreaks and estimate the effective reproduction number, with no need for explicit assumptions about the shape of epidemic growth. We demonstrate this phenomenologic approach using analytical results and simulations from mechanistic models, and provide validation against a range of empirical disease datasets. Our results suggest that sub-exponential growth in the early phase of an epidemic is the rule rather the exception. Mechanistic simulations show that slight modifications to the classical SIR model, result in sub-exponential growth, and in turn a rapid decline in the reproduction number within 3-5 disease generations. For empirical outbreaks, the generalized-growth model consistently outperforms the exponential model for a variety of directly and indirectly transmitted diseases datasets (pandemic influenza, measles, smallpox, bubonic plague, cholera, foot-and-mouth disease, HIV/AIDS, and Ebola) with model estimates supporting sub-exponential growth dynamics. The rapid decline in effective reproduction number predicted by analytical results and observed in real and synthetic datasets within 3-5 disease generations contrasts with the expectation of invariant reproduction number in epidemics obeying exponential growth. The generalized-growth concept also provides a compelling argument for the unexpected extinction of certain emerging disease outbreaks during the early ascending phase. Overall, our approach promotes a more reliable and data-driven characterization of the early epidemic phase, which is important for accurate estimation of the reproduction number and prediction of disease impact.

**Keywords:** Effective reproduction number; phenomenological model; generalized-growth model; epidemics; exponential growth; sub-exponential growth.

**Introduction**

There is a long and successful history of using compartmental transmission models to study epidemic dynamics, often calibrated using time series data describing the progression of the epidemic [1-6]. A fundamental tenet of the classic epidemic theory is that the initial growth phase should be exponential in the absence of susceptible depletion or interventions measures. However, early sub-exponential (e.g., polynomial) growth patterns have been observed in outbreaks of HIV/AIDS [7-9], Ebola [10], and foot-and-mouth disease [11]. Potential mechanisms remain debated but include spatial heterogeneity, perhaps mediated by the route of transmission (i.e., airborne vs. close contact) [10-12], clustering of contacts [8], and reactive population behavioral changes that can gradually mitigate the transmission rate [10, 11]. Accordingly, a range of mechanistic models can reproduce sub-exponential growth dynamics before susceptible depletion sets in, including models with gradually declining contact rate over time [13] and spatially-structured models such as household-community networks [12], regular lattice static contact networks, or small-world networks with weak global coupling [13, 14]. For real epidemics however, the underlying mechanisms governing sub-exponential growth can be difficult to disentangle and hence to model [13].

Given the relatively common occurrence of sub-exponential growth dynamics in empirical data and the variety of mechanisms at play, a flexible phenomenological model has been proposed to reproduce a variety of growth profiles. In the Generalized Growth Model (GGM), a tuning parameter (called the deceleration of growth, $p$) can reproduce a range of dynamics from constant incidence ($p=0$) to exponential growth ($p=1$) [11]. Application of this generalized-growth model to empirical data supports a notably slow spread ($p<1$) of the 2014 Ebola outbreaks at district-level in parts of West Africa, intermediate spread profiles for historical plague and smallpox outbreaks ($p=0.8$), and near exponential dynamics for pandemic influenza ($p \approx 1$) [15]. Departure from standard epidemic theory may be more common than previously thought since transmission heterogeneities are the rule rather than the exception [11]. In this paper, we build on the generalized-growth model concept [11, 13] to show that faithful characterization of such departures is important for accurate estimation of the reproduction number.

The basic reproduction number, commonly denoted by $R_0$, is a key parameter that characterizes the early epidemic spread in a fully susceptible population, and can be used to inform public health authorities on the level of risk posed by an infectious disease and the potential effects of intervention strategies [16]. According to the classical theory of epidemics, largely based on compartmental modeling (e.g., [1, 2, 17, 18]), $R_0$ is expected to remain invariant during the early phase of an epidemic that grows exponentially and as long as susceptible depletion remains negligible [2]. More generally, temporal variation in the transmission potential of infectious diseases are monitored via the effective reproduction number, $R_t$, defined as the average number of secondary cases per primary case at calendar time $t$ [19]. If $R_t < 1$ then the epidemic declines while $R_t > 1$ indicates widespread transmission.

Here, we expand on the generalized-growth method [11] to characterize and estimate the effective reproduction number $R_t$ during the early growth phase and before susceptible depletion sets in. We illustrate our phenomenologic approach using a combination of analytical results and synthetic datasets derived from mechanistic models with spatial or temporal effects yielding sub-exponential growth, and apply the approach to empirical data reflecting a variety historic and contemporary outbreaks. We show that sub-exponential growth dynamics is both common and important for accurate assessment of the early transmission potential.

**Materials and Methods**

Our study is organized in four sections. First, we describe the phenomenological 'generalized-growth' model and the method to estimate the effective reproduction number in this model. Second, we simulate from this phenomenologic model to gauge the expected magnitude of temporal variation in reproduction number and test our analytic predictions.

Third we simulate epidemic datasets based on mechanistic transmission models that reproduce sub-exponential growth dynamics, and apply the generalized-growth estimation approach to these synthetic data. Fourth, we test a range of empirical outbreak datasets for presence of sub-exponential growth and estimation of the effective reproduction number.

***The effective reproduction number during the early epidemic growth phase***

We extend a previously established generalized-growth model (GGM) [11] to estimate the effective reproduction number $R_g$ according to disease generations $g$. Briefly, the GGM is a useful phenomenological model that relaxes the assumption of exponential growth in the early ascending phase of an outbreak, taking the form:

$$C' = rC^p \tag{1}$$

where $C'(t)$ describes the incidence at time $t$, the solution $C(t)$ describes the cumulative number of cases at time $t$, $r$ is a positive parameter denoting the growth rate (with units of (people)$^{1-p}$ per time), and $p \in [0,1]$ is a 'deceleration of growth' parameter (dimensionless). If $p = 0$, this equation describes constant incidence over time and the cumulative number of cases grows linearly, whereas $p = 1$ describes exponential growth in the Malthus equation and the solution is given by $C(t) = C_0 e^{rt}$, where $C_0$ is the initial number of cases. An equivalent approach to model sub-exponential growth would be to modulate the growth rate rather than the cumulative number of cases (see Supplementary Information).

For the exponential growth model, the average number of secondary cases generated by initial cases during the first generation interval $T_g$ (assumed to be fixed) is estimated by [20, 21]:

$$R_0^{\exp} = \frac{C'(T_g)}{C'(0)} = \frac{rC_0 e^{rT_g}}{rC_0} = e^{rT_g} \tag{2}$$

The expression for $R_0^{exp}$ depends only on $r$ and $T_g$. Moreover, during the exponential growth phase, $R_0^{exp}$ remains invariant at $e^{rT_g}$. This can be shown by analyzing $R_g^{exp}$, the ratio of case incidences over consecutive generation intervals, which is given by:

$$R_g^{exp} = \frac{C'[(g+1)T_g]}{C'[gT_g]} = e^{r(g+1)T_g - rgT_g} = e^{rT_g} \tag{3}$$

In the case of sub-exponential growth, i.e. when $p < 1$, we can characterize the effective reproduction number $R_g^{subexp}$ over disease generations, $g$. For such polynomial epidemics, equation (1) exhibits an explicit solution that describes the cumulative number of cases over time, $C_{subexp}(t)$, in the form of [22]:

$$C_{subexp}(t) = \left[r(1-p)t + A\right]^{1/(1-p)} \tag{4}$$

where $A = C_0^{1-p}$. Hence, the corresponding incidence equation is given by:

$$C'_{subexp}(t) = r\left[r(1-p)t + A\right]^{p/(1-p)} \tag{5}$$

The analytic expression for the effective reproduction number by disease generation, $R_g^{subexp}$, captures the ratio of case incidences over consecutive disease generations:

$$R_g^{subexp} = \frac{C'_{subexp}[(g+1)T_{g+1}]}{C'_{subexp}[gT_g]} = \left[1 + \frac{r(1-p)T_g}{r(1-p)gT_g + A}\right]^{p/(1-p)} \tag{6}$$

In contrast to the exponential growth model ($p=1$) where $R_g^{\exp}$ was independent of disease generation throughout the early growth phase, we observe that $R_g^{sub\exp}$ in generation $g$ varies as a function of $g$. Given that $A$ is fixed in Equation (6), the ratio $\dfrac{r(1-p)T_g}{r(1-p)gT_g + A}$ declines to zero as $g$ increases, and thus $R_g^{sub\exp}$ approaches 1.0 asymptotically. Moreover, $R_g^{sub\exp} \to e^{rT_g}$ as $p \to 1^-$ (see Supplementary Information).

*Numerical estimation of the effective reproduction number*

The effective reproduction number can be estimated from case incidence data simulated from the generalized-growth model and using information about the distribution of the disease generation interval (Table 1). Specifically, based on the incidence at calendar time $t_i$ denoted by $I_i$, and the discretized probability distribution of the generation interval denoted by $\rho_i$, the effective reproduction number can be estimated using the renewal equation [19, 23]:

$$R_{t_i} = \dfrac{I_i}{\sum_{j=0}^{i} I_{i-j}\rho_j} \quad (7)$$

where the denominator represents the total number of cases that contribute (as primary cases) to generating the number of new cases $I_i$ (as secondary cases) at calendar time $t_i$ [19].

***Trends in effective reproduction number based on simulations from the phenomenological 'generalized growth' model***

To gauge the expected temporal variation in the effective reproduction number for a range of growth profiles and test analytic predictions, we simulate incidence data using the generalized-growth model (Equation 1). We fix the growth rate parameter $r$, but assume different distributions of the disease generation interval (e.g., exponential, gamma, uniform, delta), and vary the 'deceleration of growth' parameter $p$ between 0 and 1 [11]. We analyze the outbreak trajectory in the first 5 disease generations to estimate the effective reproduction number using Equation (7) (assuming a fixed generation interval) and compare with estimates obtained from the analytic expressions in Equations 3 and 6.

*Trends in early growth dynamics based on mechanistic models simulations*

Next, we develop three specific examples of mechanistic transmission models that support early sub-exponential growth dynamics. These include: (1) SIR (susceptible-infectious-removed) dynamics on a spatially structured model, as substantial levels of clustering have been hypothesized to yield early polynomial epidemic growth [8, 11, 12] (2) SIR compartmental model with reactive behavioural changes *via* a time-dependent transmission rate [10, 11, 24-26]; and (3) an SIR compartmental model with inhomogeneous mixing [27, 28]. We briefly describe these models below.

<u>SIR epidemics on a spatially structured model</u>

One of the putative mechanisms leading to early polynomial epidemic growth dynamics is clustering [8, 11], a network property that quantifies the extent to which the contacts of one individual are also contacts of each other [14]. Contact networks are particularly useful to explore the impact of clustering; here we use a network-based transmission model with household-community structure, which has been previously applied to study the transmission dynamics of Ebola [12, 29]. In this model, individuals are organized within households of size $H$ (each household contains $H$ individuals) and households are organized within communities of size $\bar{C}$ households (each community contains $\bar{C} \times H$ individuals). Network connectivity is identical for every individual. The household reproduction number $R_{0H}$ was set at 2.0 and the community reproduction number $R_{0c}$ was set at 0.7 based on previous study [12]. For a fixed household size ($H = 5$) and different values of the community size parameter, we analyze the temporal profile in case incidence and the effective reproduction number during the first few disease generations from 200 independent stochastic realizations.

*SIR compartmental model with reactive behavioural changes*

In addition to contact clustering, rapid onset of behavior changes is another mechanism that has been hypothesized to lead to sub-exponential growth dynamics, as it would result in an early decline in effective reproduction number. For instance, during the 2014-15 Ebola epidemic, some areas of West Africa exhibited early sub-exponential growth even before control interventions were put in place [10, 30].

To model behavior change, we consider a classical SIR epidemic model [1, 3] with time-dependent contact rate following:

$$\begin{aligned}
\frac{dS(t)}{dt} &= -\beta(t) S \frac{I}{N} \\
\frac{dI(t)}{dt} &= \beta(t) S \frac{I}{N} - \gamma I \\
\frac{dR(t)}{dt} &= \gamma I
\end{aligned} \quad (8)$$

where $S(t)$, $I(t)$, and $R(t)$ denote the number of susceptible, infectious and removed (recovered) hosts in a randomly-mixed population of size $N$, $\beta(t)$ is the time-dependent transmission rate, the probability that a susceptible individual encounters an infectious individual is given by $I(t)/N$, and $1/\gamma$ is the mean infectious period.

In the classical SIR model with constant transmission rate $\beta$, in a completely susceptible population, $S(0) \approx N$, and $I(t)$ grows exponentially during the early epidemic phase, e.g., $I(t) \approx I_0 e^{\gamma(R_0-1)t}$, where $R_0 = \beta/\gamma$ is the average number of secondary cases generated by a primary case during the infectious period. When susceptible depletion kicks in ($S(t)<S(0)$), the effective reproduction number, $R_t$, declines following $R_t = \frac{S(t)}{N} R_0$. During the first few disease generations, where $S(0)/N \approx 1$, the classical SIR model supports a reproduction number that is nearly invariant, i.e., $R_t \approx R_0$. Here, to capture behavior change, we model an exponential decline in the transmission rate $\beta(t)$ from an initial value $\beta_0$ towards $\phi\beta_0$ at rate $q > 0$ following:

$$\beta(t) = \beta_0\left((1-\phi)e^{-qt} + \phi\right)$$

Here $\beta(t)$ leads to early sub-exponential growth dynamics whenever $R_0 > 1$ and $q > 0$. Assuming that $R_0 > 1$ in a sufficiently large susceptible population so that the effect of susceptible depletion is negligible in the early epidemic phase, the quantity $1-\phi$ models the proportionate reduction in $\beta_0$ that is needed for the effective reproduction number to asymptotically reach 1.0. Hence, $\phi$ can be estimated as $1/R_0$. If $q = 0$, $\beta(t) = \beta_0$ and we recover the classic SIR transmission model with early exponential growth dynamics. In general, a faster decline of the effective reproduction number towards 1.0 occurs for higher values of $q$, even without susceptible depletion. It is worth noting that prior HIV/AIDS models (e.g., [31]) have incorporated exponential decay in the transmission rate in a similar manner as described here, albeit the rate of decay was assumed to be a time-dependent function of HIV/AIDS prevalence.

To examine the behaviour of the effective reproduction number $R_g$ over disease generations in the above model, we analyze the temporal progression in the number of cases at generation $g$ based on the following discrete equations [32]:

$$I_{g+1} = S_g(1 - e^{-R_0\left((1-\phi)e^{-qg} + \phi\right)\frac{I_g}{N}}) \quad (9)$$
$$S_{g+1} = S_g - I_{g+1}$$

where $\phi = \frac{1}{R_0}$, $I_g$ is the number of new cases at generation $g$ and $S_g$ is the number of remaining susceptibles at generation $g$. We initialize simulations with $I_0 = 1$ and $S_0 = N$ where $N$ is set to $10^8$ individuals.

*SIR compartmental model with nonlinear incidences*

Beyond contact clustering and decay in transmission rate, a third mechanism potentially accounting for sub-exponential growth is departure from mass action, which may arise due to spatial structures or other forms of nonhomogenous mixing. These effects can be incorporated in SIR models using nonlinear incidence rates (see e.g., [33, 34]). For instance, the incidence rate can take the form: $\beta_0 S(t) I^\alpha(t)/N$ where $\alpha$ is a phenomenological scaling mixing parameter; $\alpha = 1$ models homogeneous mixing while $\alpha < 1$ reflects contact patterns that deviate from random mixing and lead to slower epidemic growth [35]. A related version of this model is the TSIR model [28], which has found applications in various infectious disease systems including measles [28, 36], rubella [37], and dengue [38].

Here we consider an SIR model with non-homogeneous mixing, with constant transmission rate β₀ and mixing parameter α, following:

$$\frac{dS(t)}{dt} = -\beta_0 S \frac{I^\alpha}{N}$$
$$\frac{dI(t)}{dt} = \beta_0 S \frac{I^\alpha}{N} - \gamma I \qquad (10)$$
$$\frac{dR(t)}{dt} = \gamma I$$

To analyze the progression of the reproduction number $R_g$ over disease generations in the above model, we use the following discrete equations describing the number of cases at generation $g$ [13, 32]:

$$I_{g+1} = S_g(1 - e^{-R_0 \frac{I_g^\alpha}{N}})  \quad (11)$$
$$S_{g+1} = S_g - I_{g+1}$$

where $I_g$ is the number of new cases at generation $g$ and $S_g$ is the number of remaining susceptible individuals at generation $g$. We initialize simulations with $I_0 = 1$ and $S_0 = N$ where $N$ is set to $10^8$ individuals.

*Application to real outbreak data*

Lastly, we analyze a variety of empirical outbreak datasets to test the importance of sub-exponential growth in observed disease dynamics and the resulting impact on the effective reproduction number estimates. We rely on a convenience sample representing a variety of pathogens, geographic contexts, and time periods, and include outbreaks of pandemic influenza, measles, smallpox, bubonic plague, cholera, foot-and-mouth disease (FMD), HIV/AIDS, and Ebola (Table 1 and Supplementary Information for time series). The temporal resolution of the datasets varies from daily to annual. For each outbreak, the onset week corresponds to the first observation associated with a monotonic increase in the case incidence, up to the peak incidence.

We focus on the first 3-5 disease generations, depending on the length of the available empirical time series. We estimate the effective reproduction number using the 2-step approach. In the first step, we use nonlinear least-squares to fit the generalized growth model to the synthetic mechanistic data, and estimate parameters $r$ and $p$ (Equation 1, [11]). The initial number of cases $C_0$ is fixed according to the first observation. Nominal 95% confidence intervals for parameter estimates $r$ and $p$ are constructed by simulations of 200 best-fit curves $C'(t)$ using parametric bootstrap with a Poisson error structure, as in prior studies [39]. In the second step, we simulate epidemic curves using the generalized-growth model with estimated *r* and *p*, and apply equation (7) to the simulated incidence data. We assume a gamma distribution for the generation interval, with means and standard deviations as in Table 1 [40-46]. Also, for each outbreak, we compare the goodness of fit of the phenomenological generalized-growth model vs. the exponential growth models.

## Results

*Trends in effective reproduction number based on simulations from the phenomenological 'generalized growth' model*

We first analyze simulations of epidemic growth under the generalized growth model in the first 5 disease generations of the outbreak, for different values of *r* and *p* and a fixed generation interval (Figure 1 and Figure S2). Our simulations confirm the analytical results described in Equations (3) and (6) in relation to changes in the effective reproduction number under early exponential ($p=1$) and sub-exponential growth dynamics ($p<1$). As expected, the greater the departure from exponential growth (*p* close to 0), the lower the effective reproduction number $R_g^{sub\exp}$. But more importantly, in the case of sub-exponential growth, and for a given growth rate $r$, the effective reproduction number $R_g^{sub\exp}$ is a dynamic quantity that approaches 1.0 asymptotically with increasing disease generations. In contrast, for exponential growth ($p=1$), the effective reproduction number remains invariant during the early epidemic growth phase.

We also run simulations under different assumptions regarding the distribution of the generation interval and vary $p$ in the range $0<p\leq1$ (Figure 2). The declining trend in the effective reproduction number associated with the sub-exponential growth regime ($p<1$) persists independently of the generation interval distribution. Moreover, as $p$ decreases, estimates of the effective reproduction number become less dependent on the generation interval distribution (Figure 2). This indicates that for a sufficiently small $p < 1$, the mean of the generation interval distribution provides sufficient information to estimate the reproduction number, without the need to specify a full distribution.

## Trends in case incidence and the effective reproduction number based on mechanistic models simulations

### SIR epidemics on a spatially structured epidemic model

Figure 3 shows simulations of case incidence and the effective reproduction number $R_g$ derived from the household-community transmission model for different levels of community mixing, tuned by $\bar{C}$. As expected, the lower the community mixing, the greater the departure from homogeneous mixing, and hence the greater the departure from early exponential growth dynamics. Early sub-exponential growth dynamics are observed in all community mixing scenarios tested ($\bar{C}$=25, 45 and 65 households), which is consistent with a declining trend in the effective reproduction number $R_g$ (Figure 3).

### SIR compartmental model with reactive behavioural changes

Representative profiles of $R_g$ for the SIR model with time-dependent transmission rate $\beta(t)$ are shown in Figure 4 for different values of the speed of transmission decline, tuned by parameter $q$. The decline in effective reproduction number $R_g = I_{g+1}/I_g$ ($g = 0 \ldots n$) is more pronounced as the decline in transmission rate is faster (i.e, $q \gg 0$). Early sub-exponential growth dynamics is seen in all simulations where $q > 0$ (Figure 4).

### SIR compartmental model with nonlinear incidence rates

Simulations for this model display concave down incidence curves in semi-logarithmic scale, supporting the presence of early sub-exponential growth, even for values of $\alpha$ slightly below the homogenous mixing regime (i.e., α just below 1, Figure S3). Accordingly, the effective reproduction number $R_g$ exhibits a declining trend during the first few disease generations (Figure 5). By contrast, the reproduction number remains invariant at $R_g = R_0 = 2$ when $\alpha = 1$ (Figure 5(B), black curve).

*Application to real outbreak data*

Lastly, we apply the concepts of sub-exponential growth dynamics to a variety of empirical outbreak datasets. Figures S4-S5 provide a comparative analysis of the goodness of fit provided by the generalized-growth and the exponential growth models across outbreaks. Our results indicate that the generalized-growth model consistently outperforms the exponential growth model in the early ascending phase of the outbreak, even when p is only slightly below 1.0 (i.e., departure from the exponential model is slight). Across outbreaks, we find variability in the deceleration of growth parameter, even for a given pathogen (median $p = 0.57$, IQR: 0.46–0.84; Figure 6). Not surprisingly, parameter uncertainty declines with increasing length of the early epidemic phase used for estimation (Figures 6-7). On the other hand, mean estimates of $p$ (Table 1) are stable during the first 3-5 disease generations (ANOVA, P=0.9).

When we use the generalized-growth model to estimate the effective reproduction number, we find a declining trend in the effective reproduction number with increasing disease generation intervals and variability in estimates of the reproduction number across 21 outbreaks representing 8 different pathogens (Figure 7). Further, estimates of the effective reproduction number are sensitive to small changes in the deceleration of growth parameter across outbreaks (Spearman rho>0.62, P<0.002; Figure S6).

Model fits to empirical data illustrates a variety of exponential and sub-exponential growth profiles across pathogens (Figures S7-S11). For instance, the autumn 1918 influenza pandemic in San Francisco is characterized by near exponential growth, with $p \sim 0.8 - 0.9$ and a relatively stable reproduction number in the range 1.7–1.8 (Figure S7). In contrast, the foot-and-mouth disease outbreak in Uruguay at the farm level displays slower initial growth with mean $p \sim 0.4 - 0.5$ and a more variable reproduction number in the range 1.6–2.8 (Figure S8). For the HIV/AIDS epidemic in Japan (1985-2012), we estimated the mean effective reproduction number in the range 1.3–1.6 with $p \sim 0.5$ assuming a mean generation interval of 4 years, consistent with pronounced departure from exponential growth (Figure S9).

The wealth of district-level Ebola data available for the 2014 epidemic in West Africa provides a good opportunity to gauge geographic variations in the growth profiles, and in the resulting effective reproduction numbers. Indeed, we find variability across geographic locations in the effective reproduction number (median=1.46, IQR: 1.26–1.83) and deceleration parameter $p$ (median=0.58, IQR: 0.46–0.72), and correlation between these parameter estimates (Spearman rho=0.81, P<0.001). For comparison, at the $5^{th}$ disease generation interval, the highest estimate of the effective reproduction number was at 2.5 (95% CI: 2.0 – 2.7) for the 2014 Ebola outbreak in Montserrado, Liberia while the lowest estimate was at 1.03 (95% CI: 1 –1.1) for the outbreak in Bomi, Liberia (Figure 7).

**Discussion**

In this study, we introduce a quantitative 'generalized growth' framework to characterize the transmission potential of pathogens in the early phase of an outbreak, when susceptible depletion remains negligible, without making explicit assumptions about the epidemic growth profile. The phenomenological 'generalized growth' model reproduces a range of growth dynamics from polynomial to exponential [11] and is agnostic of the mechanisms affecting growth, which may include contact patterns, spatial effects, non-homogeneous mixing, and/or behavior changes. A phenomenological model can be particularly useful when biological mechanisms are difficult to identify. Using a combination of analytical results, simulations from mechanistic models, and analyses of empirical outbreak data, we demonstrate that the effective reproduction number typically displays a downward trend within the first 3-5 disease generations. Evidence of sub-exponential growth, and associated reproduction number decline, is found in disease systems as varied as Ebola, pandemic influenza, smallpox, plague, cholera, measles, foot-and-mouth disease, and HIV/AIDS. Our results indicate that the concept of sub-exponential growth is both widespread and important to consider for accurate assessment of the reproduction number.

For epidemics that truly depart from exponential growth theory, traditional estimation methods relying on the assumption of exponential growth are expected to inflate reproduction number estimates. The bias between theoretical values and estimates increases as departure from exponential growth becomes more pronounced, i.e., when $p$ decreases towards 0, representing slower epidemic spread compared to the exponential case where $p=1$. For instance, our estimate of the reproduction number for the 1972 smallpox epidemic in Khulna, Bangladesh (~2 (95% CI: 1.6–2.6)) is significantly lower than earlier historic estimates of smallpox based on an exponential growth assumption (range 3.5 – 6.0) [47]. In contrast, when $p$ is near 1.0, indicating near exponential growth, our estimates of the reproduction number remain consistent with those of compartmental models. This is the case for the 1905 bubonic plague epidemic in Bombay, India [48] or the 1918 influenza pandemic in San Francisco [15]. Overall, our estimates for Ebola outbreaks tend to be slightly lower than those reported in prior studies, possibly because of sub-exponential growth at the district levels [12, 26, 43, 49-54]. It is also worth noting that the incorporation of generalized-growth in a phenomenological logistic-type model can substantially increase the performance of the model for short-term forecasting and prediction of the final epidemic size as recently illustrated in the context of the Zika epidemic in Colombia [55].

Here we have studied simulations from three common types of mechanistic models supporting early sub-exponential growth dynamics and incorporating characteristics of the host contact network, behavior changes, and inhomogeneous mixing. In all models, relatively small departures from crude SIR dynamics led to sub-exponential growth profiles, and in turn a quick decline in effective reproduction number, speaking to the generalizability of our findings. With real outbreak case series data however, it can be difficult to disentangle the mechanism or combination of mechanisms shaping the early epidemic growth profile, especially when case series data are limited to the early epidemic growth phase. In order to assess the contribution of different mechanisms, independent sources of data would be required to quantify the structural characteristics of the contact network [56-58], or the timing and intensity of possible behavior changes. Further, for comparison purposes, it can be particularly useful to enumerate secondary cases from transmission tree data whenever available, and obtain independent estimates of R0 [56-58] agnostic of any model form. There is clearly scope for more research work in this area. In the absence of detailed information on the chains of transmission or the biological mechanisms at play however, a phenomenological approach such as that proposed here with the generalized-growth model may be preferable.

In addition to providing a quantitative framework for estimation of the reproduction number based on a phenomenological approach, this study has implications for disease control, particularly our understanding of herd immunity and extinction thresholds [1, 2]. In the simple SIR models, the critical fraction of the population needed to be effectively vaccinated to prevent an epidemic is given by $1-1/R_0$, which is in the range 50–90% of the population for most epidemic diseases [1, 54]. However, this fraction may be potentially considerably lower for epidemics rendering sub-exponential growth, where the effective reproduction number naturally declines towards unity irrespective of other intervention measures and before susceptible depletion sets in. For example, the 2014 West-African Ebola outbreak ended with less than 1% of the population registered as cases, which defies expectations from SIR models, and the contribution of large-scale interventions on these low attack rates remains debated [59]. These data-driven observations suggest that more attention should be paid to the shape of the early ascending phase of emerging infectious diseases outbreaks, and the associated uncertainty in the reproduction number estimates should be considered.

A related consequence of sub-exponential growth dynamics, and associated decline in effective reproduction number, is the effect on the extinction threshold. Indeed, it is natural to expect a higher probability of extinction due to stochastic effects for epidemics governed by sub-exponential growth. This may in part explain the small magnitude and a short duration of most Ebola outbreaks since 1976 [60-62], as Ebola showed substantial departure from exponential growth ($0.6<p<0.72$). In fact, simulations using an individual-level stochastic model for Ebola with household and community contact network structure are consistent with early sub-exponential growth dynamics, and has a probability of ~40% of spontaneous die-out of an outbreak within the first month of transmission [12]. This model [12] is also consistent with an effective reproduction number that asymptotically declines toward unity as the virus spreads through the population. From a public health perspective, outbreaks characterized by sub-exponential growth dynamics may provide a greater window of opportunity for implementation of control interventions compared to those following exponential or near exponential growth dynamics [12].

Overall, our results underscore the need to carefully characterize the shape of the epidemic growth phase in order to accurately assess early trends in reproduction number. Consideration of the sub-exponential growth phenomenon will improve our ability to appropriately model transmission scenarios, assess the potential effects of control interventions, and provide accurate forecasts of epidemic impact. Looking to the future, the development of new mechanistic transmission models is needed to provide a better understanding of the factors shaping early epidemic growth. Such models would allow for systematic evaluation of epidemic outcomes and disease control policies. A recent review of forecasting models for the West-African Ebola epidemic highlighted a range of approaches to investigating disease spread from simple phenomenological models, to compartmental epidemic models, to intricate contact networks [63]. The vast majority of these approaches considered early exponential growth dynamics, an assumption that led to substantial overestimation of Ebola epidemic size and peak timing and intensity. In light of these findings, Chretien et al. [63] stress the need for new mechanistic models that incorporate "dampening approaches" to improve characterization of the force of infection and provide uniform forecasting approaches and evaluation metrics. We believe the present study represents a significant step in this direction.

**Data Accessibility**

All of the epidemic incidence data employed in this paper is being made publicly available in a supplementary file.

**Competing interests**

We have no competing interests.

**Authors' contributions**

GC designed the study. GC and SM contributed to methods. GC carried out simulations, analyzed the data, and wrote the first draft of the manuscript. GC, CV, LS and SM contributed to the writing and revisions of the manuscript, and approved its final version.

**Funding**

GC acknowledges financial support from the NSF grant 1414374 as part of the joint NSF-NIH-USDA Ecology and Evolution of Infectious Diseases program; UK Biotechnology and Biological Sciences Research Council grant BB/M008894/1, NSF-IIS RAPID award #1518939, and NSF grant 1318788 III: Small: Data Management for Real-Time Data Driven Epidemic simulation, and the Division of International Epidemiology and Population Studies, The Fogarty International Center, US National Institutes of Health. CV and LS acknowledges financial support from the RAPIDD Program of the Science & Technology Directorate and the Division of International Epidemiology and Population Studies, The Fogarty International Center, US National Institutes of Health. LS also acknowledges generous support from a EC Marie Curie Horizon 2020 fellowship. SM acknowledges the support from the Natural Sciences and Engineering Research Council of Canada (NSERC), and the Mathematics of Information Technology and Complex Systems (Mitacs).

**Table 1**. Summary of empirical disease datasets and our corresponding estimates of the reproduction number using 4 disease generations derived using the methodology described in the main text.

| Disease | Outbreak | Disease generation interval mean (S.D) [&] | Time series temporal resolution | Epidemic peak timing (no. data points) | Early epidemic phase (no. data points) | Reproduction number (95% CI) | Data Source |
|---|---|---|---|---|---|---|---|
| Pandemic influenza | San Francisco (1918) | 3 (1) | Days | 33 | 14 | 1.67 (1.43,2.08) | [15] |
| Smallpox | Khulna, Bangladesh (1972) | 14 (2) | Weeks | 10 | 7 | 1.97 (1.61,2.62) | [64] |
| Plague | Bombay (1905-06) | 7 (2) | Weeks | 19 | 7 | 1.37 (1.19,1.61) | [65] |
| Measles | London (1948) | 14 (2) | Weeks | 16 | 8 | 1.31 (1.26,1.36) | [66] |
| Cholera | Aalborg (1853) | 5 (1) | Days | 39 | 20 | 1.87 (1.64,2.23) | [67] |
| FMD | Uruguay (2001) | 5(1) | Days | 33 | 10 | 1.77 (1.33,2.44) | [68, 69] |
| HIV/AIDS | Japan (1985-2012) | 4 (1.4) | Years | 24 | 16 | 1.40 (1.37,1.43) | [70] |
|  | NYC (1982-2002) | 4 (1.4) | Years | 12 | 8 | 2.39 (2.37,2.44) | [71] |
| Ebola | Uganda (2000) | 13.1 (6.6) | Weeks | 8 | 7 | 1.65 (1.42,2.08) | [26, 72] |
|  | Congo (1976) | 13.1 (6.6) | Days | 22 | 16 | 3.79 (2.31,6.39) | [73, 74] |
|  | Gueckedou, Guinea (2014) | 19 (11) | Weeks | 22 | 12 | 1.46 (1.24,1.90) | [75] |
|  | Montserrado, Liberia (2014) | 13.1 (6.6) | Weeks | 16 | 8 | 1.46 (1.24,1.93) | [75] |
|  | Margibi, Liberia (2014) | 13.1 (6.6) | Weeks | 14 | 8 | 1.99 (1.52,2.86) | [75] |
|  | Bomi, Liberia (2014) | 13.1 (6.6) | Weeks | 10 | 8 | 1.05 (1.00,1.20) | [75] |
|  | Grand Bassa, Liberia (2014) | 13.1 (6.6) | Weeks | 7 | 8 | 1.25 (1.03,1.64) | [75] |
|  | Western Area Urban, Sierra Leone (2014) | 11.6 (5.6) | Weeks | 18 | 8 | 1.26 (1.16,1.41) | [75] |
|  | Western Area Rural, Sierra Leone (2014) | 11.6 (5.6) | Weeks | 19 | 8 | 1.79 (1.50,2.26) | [75] |
|  | Bo, Sierra Leone (2014) | 11.6 (5.6) | Weeks | 15 | 8 | 1.50 (1.21,1.99) | [75] |
|  | Bombali, Sierra Leone (2014) | 11.6 (5.6) | Weeks | 12 | 7 | 1.94 (1.53,2.14) | [75] |
|  | Kenema, Sierra Leone (2014) | 11.6 (5.6) | Weeks | 12 | 7 | 1.24 (1.11,1.41) | [75] |
|  | Port Loko, Sierra Leone (2014) | 11.6 (5.6) | Weeks | 16 | 7 | 1.32 (1.18,1.50) | [75] |

FMD: foot-and-mouth disease
[&]Mean and SD for generation interval are given in days except for HIV/AIDS datasets given in years.

**Figure captions**

**Figure 1**. Simulated profiles of epidemic growth (A) and corresponding effective reproduction number (B) based on the generalized-growth model during the first 5 disease generation intervals. The parameter $p$ is varied from 0.2 to 1.0 while the growth rate parameter $r$ is fixed at 0.65 per day and $C_0 = 1$. The length of the generation interval is assumed to be fixed at $T_g = 3$ days. When $p = 1$ (exponential growth), the reproduction number is constant at $R_0^{\exp} = e^{rT_g} = 7.03$. When $p < 1$ (sub-exponential growth) the reproduction number gradually declines over time and approaches 1.0 asymptotically. The black circles correspond to $R_g^{subexp}$ calculated using the analytical formula in Equation (6) while the solid lines correspond to the estimates from numerical solutions based on renewal Equation (7).

**Figure 2**. Simulated profiles of the reproduction number during the first 5 disease generation intervals under the generalized-growth model, for different values of $p$. The parameter $r$ is fixed at 0.65 per day and $C_0 = 1$. Estimates of the reproduction number are numerically obtained using Equation (7) assuming four different distributions for the generation interval: a) uniform distribution in the range 2–4 days, b) exponential distribution with the mean of 3 days, c) gamma distribution with the mean of 3 days and variance of 1 day, and d) fixed generation interval at 3 days. When $p = 1$ (exponential growth) the reproduction number is invariant irrespective of the shape of the generation interval distribution and is given by $1 + rT_g = 2.95$ when the generation interval is exponentially distributed [20], $e^{rT_g} = 7.03$ when the generation interval is fixed [20], and $(1 + rb)^a = 5.84$ when the generation interval follows a gamma distribution with the shape and scale parameters given by $a = 9$ and $b = 1/3$ [20], respectively. For $p < 1$ (sub-exponential growth), the effective reproduction number approaches 1.0 asymptotically

over disease generations. The reproduction number $R_g^{subexp}$ at fixed generation intervals can be explicitly computed using Equation (6) (black circles).

**Figure 3.** Simulations of case incidence (A, C, E) and the effective reproduction number $R_g$ (B, D, F) for the first few consecutive disease generations, using a network-based transmission model with household-community structure [12, 29] for different values of the community mixing parameter ($\bar{C}$=25, 45, and 65 households) while keeping the household size H fixed at 5. The household reproduction number $R_{0H}$ was set at 2.0 and the community reproduction number $R_{0c}$ was set at 0.7 based on a previous Ebola mathematical modeling study [12]. Each simulation started with one infectious individual. Early sub-exponential growth dynamics are observed across all scenarios, which is consistent with a declining trend in the effective reproduction number $R_g$. For comparison with exponential growth dynamics, the dark solid line illustrates exponential growth in case incidence over disease generations with a fixed reproduction number of 2.7. The horizontal dashed line at $R_g$=1.0 is shown for reference.

**Figure 4**. Simulations of case incidence (A) and the effective reproduction number $R_g$ (B) for the first few consecutive disease generations, using model (9) for different values of the decline rate parameter $q$ with $R_0 = 2$ and a large population size ($N = 10^8$). Each simulation started with one infectious individual. When $q=1$, exponential growth during the early epidemic phase is evident as a straight line fits well for several consecutive disease generations of the incidence curve in semi-logarithmic scale, and the effective reproduction number, $R_g$, remains invariant at 2. A concave down curve is indicative of sub-exponential growth with $q>0$. The horizontal dashed line at $R_g$=1.0 is shown for reference.

**Figure 5.** Simulations of case incidence (A) and the effective reproduction number $R_g$ (B) for the first few consecutive disease generations, using model (11) for different values of scaling parameter $\alpha$, with $R_0 = 2$, $\gamma = 1/5$, and a large population size ($N = 10^8$). Each simulation started with one infectious individual. When $\alpha = 1$, exponential growth during the early epidemic phase is evident as a straight line fits well for several consecutive disease generations of the incidence curve, and the effective reproduction number, $R_g$, remains invariant at 2. The horizontal dashed line at $R_g = 1.0$ is shown for reference.

**Figure 6.** Estimates of the deceleration of growth parameter and the corresponding 95% confidence intervals derived from various infectious disease outbreak datasets of case incidence series by fitting the generalized-growth model to the initial epidemic phase comprising of approximately the first 3 (green), 4 (blue), and 5 (red) disease generation intervals.

**Figure 7**. Estimates of the effective reproduction number and the corresponding 95% confidence intervals derived from various infectious disease outbreak datasets of case incidence series by fitting the generalized-growth model to the initial epidemic phase comprising of approximately the first 3 (green), 4 (blue), and 5 (red) disease generation intervals. The generation interval is assumed to follow a gamma distribution with the corresponding mean and variance provided in Table 1.

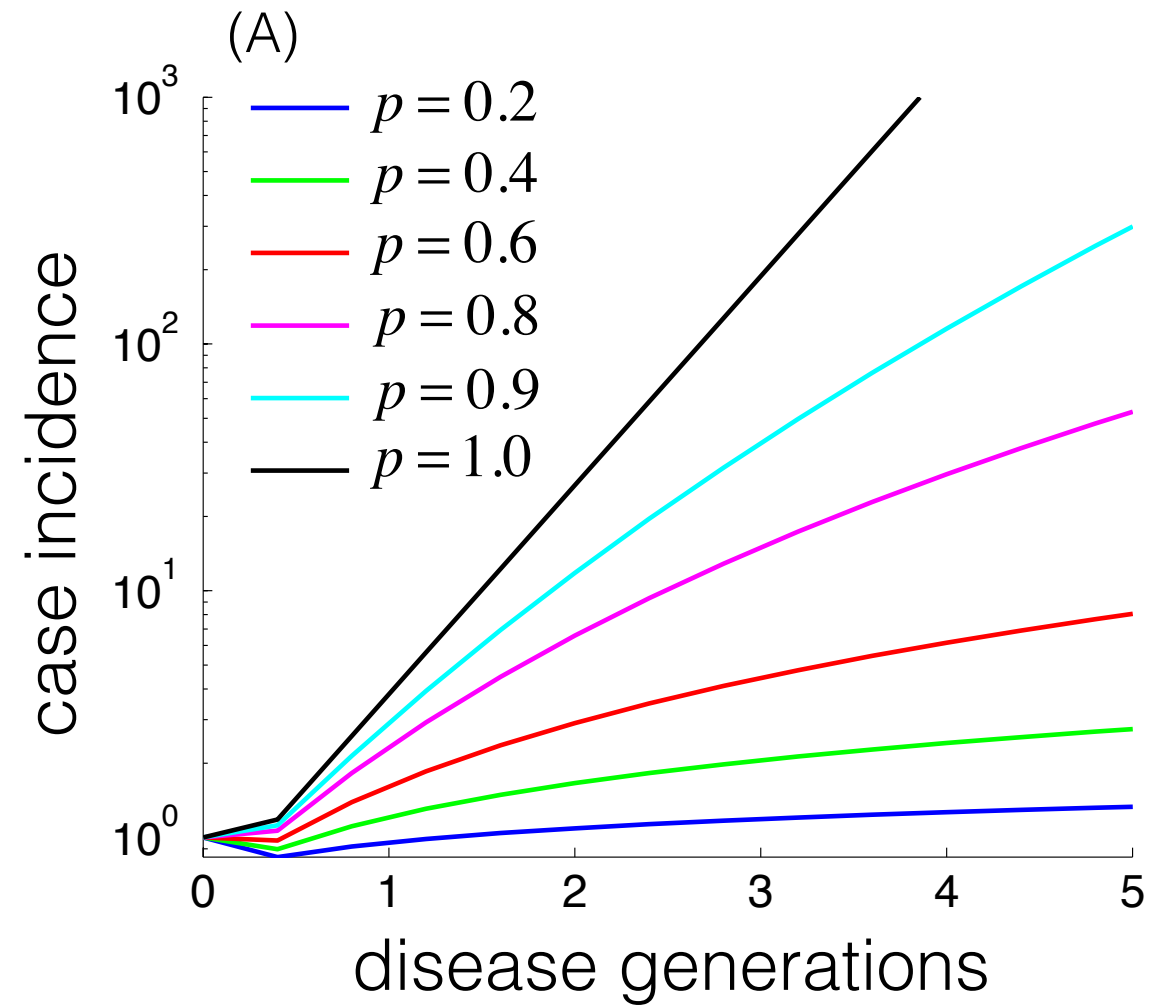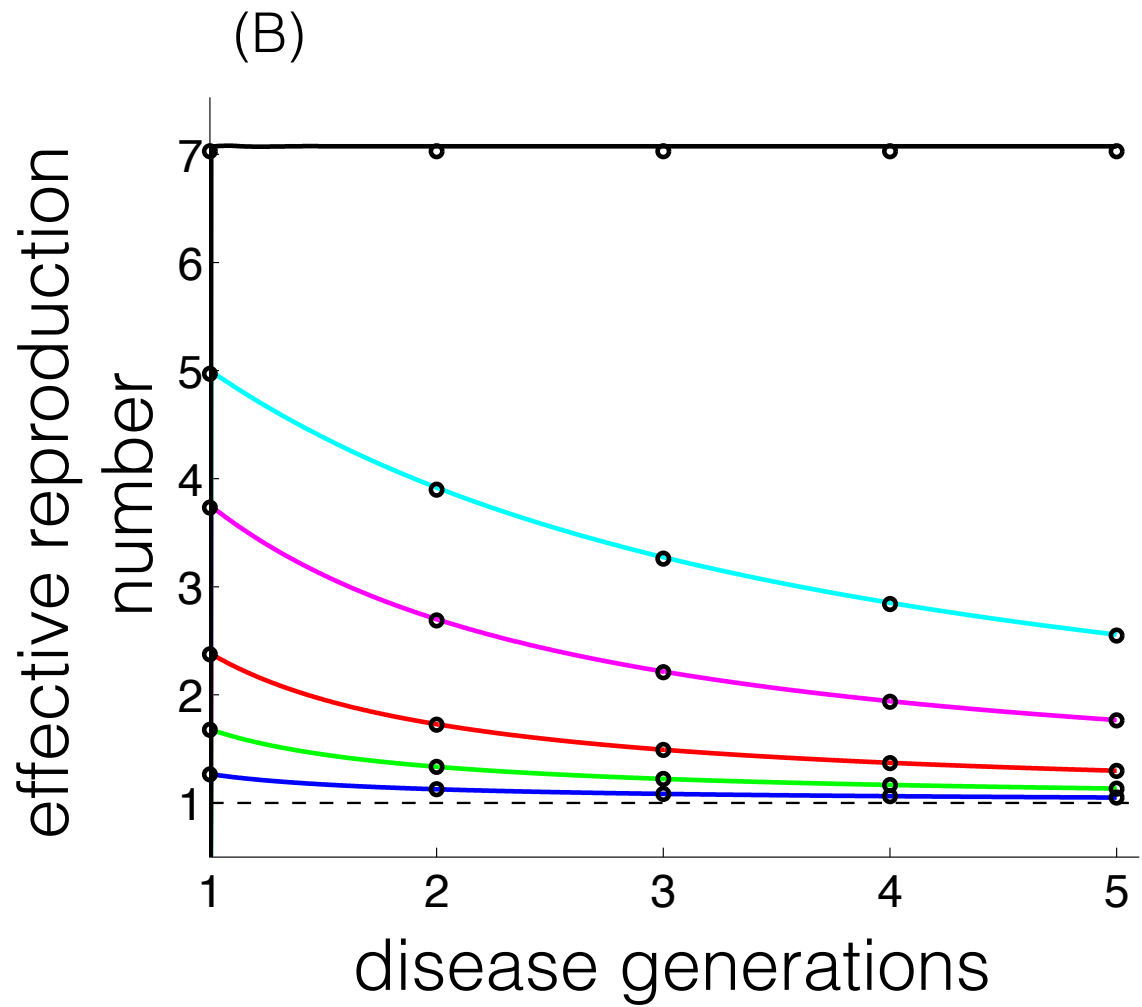

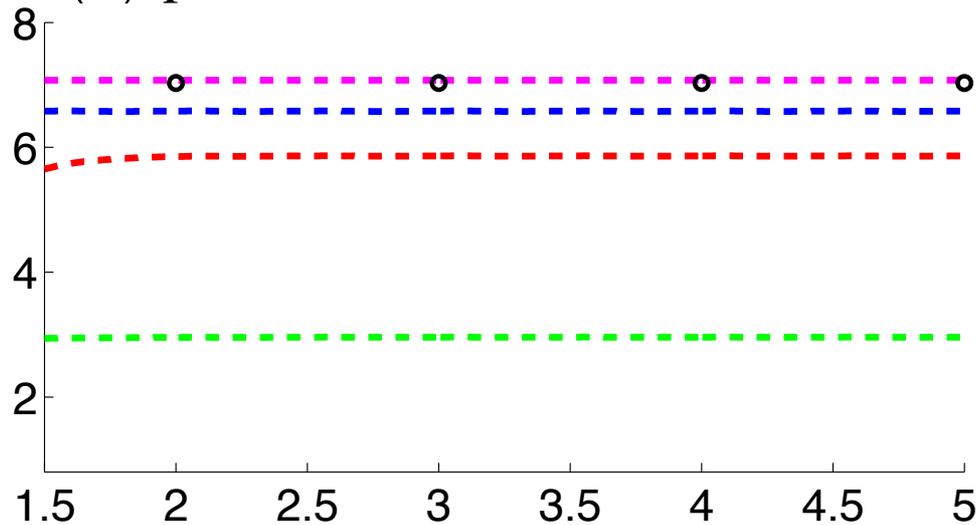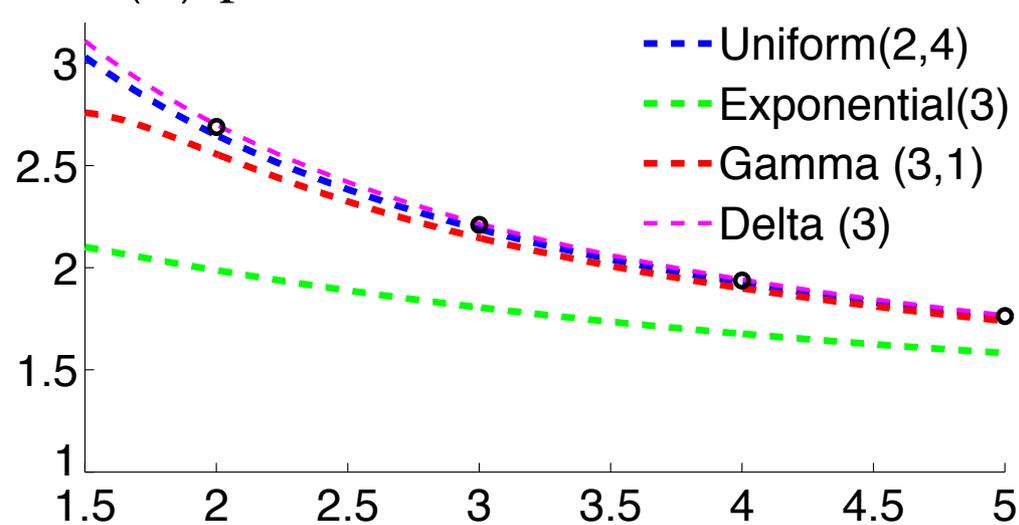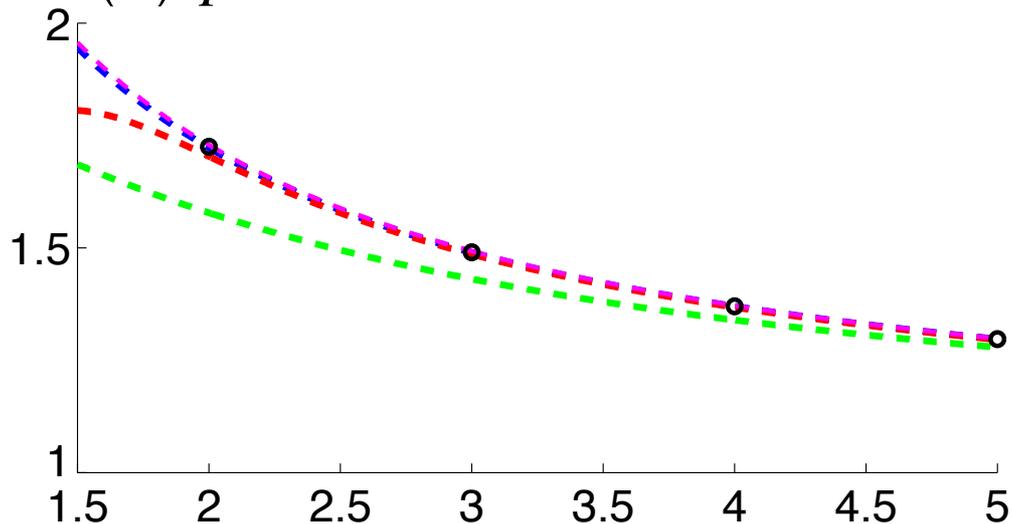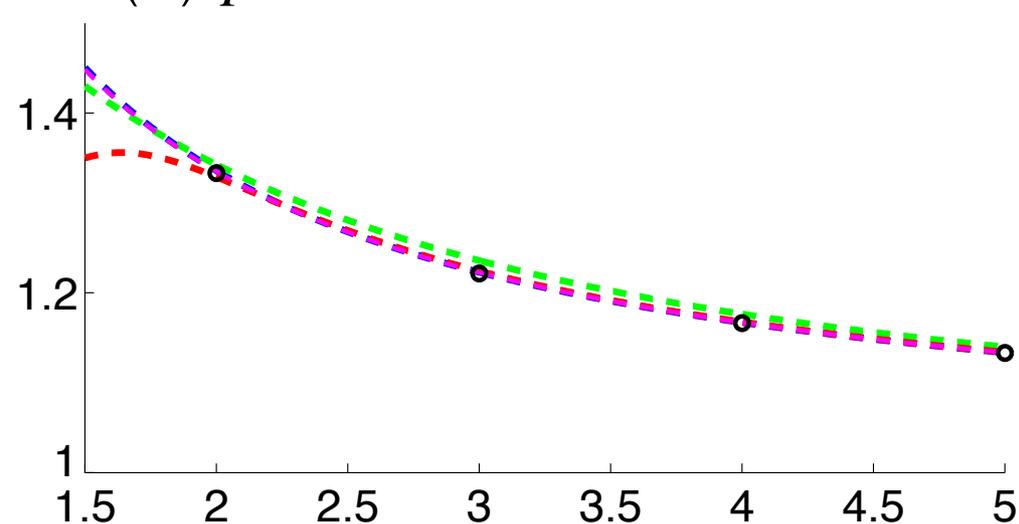

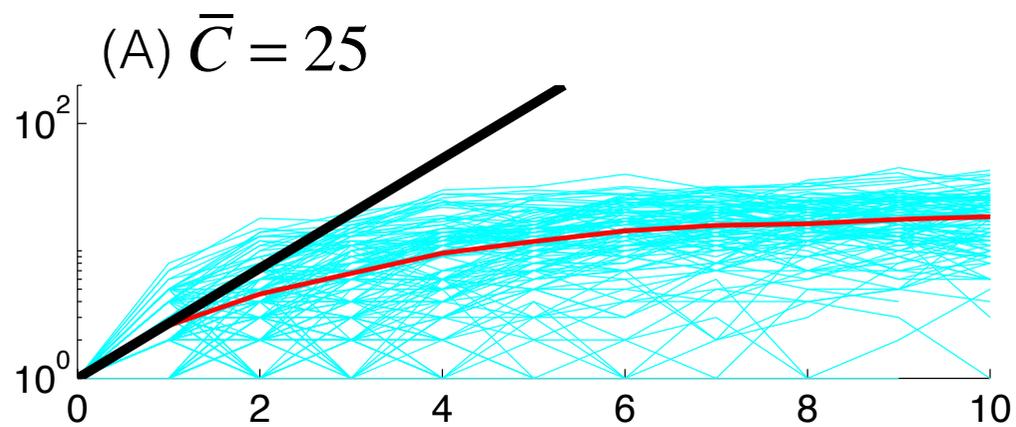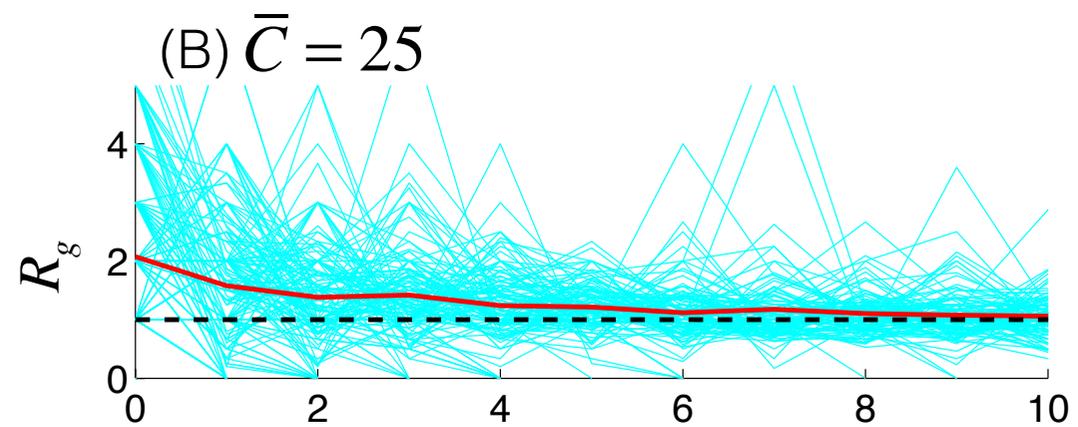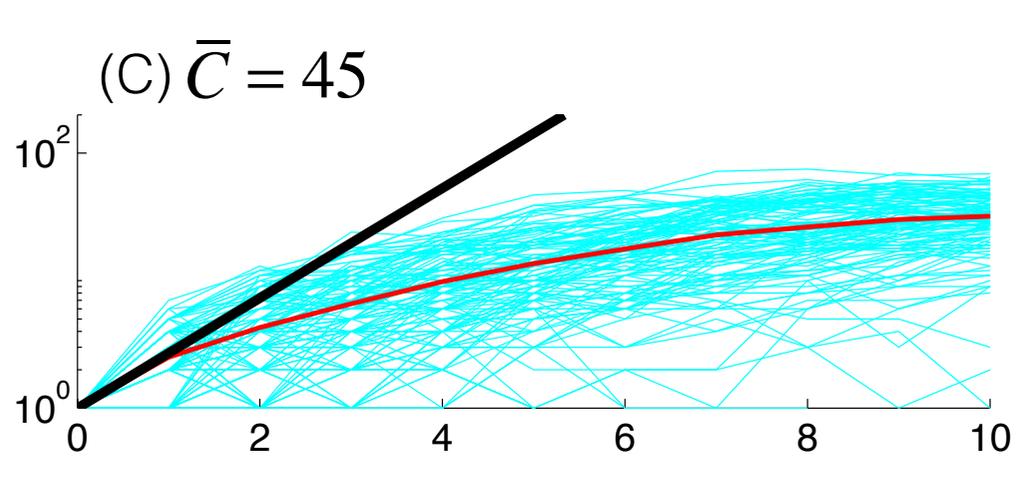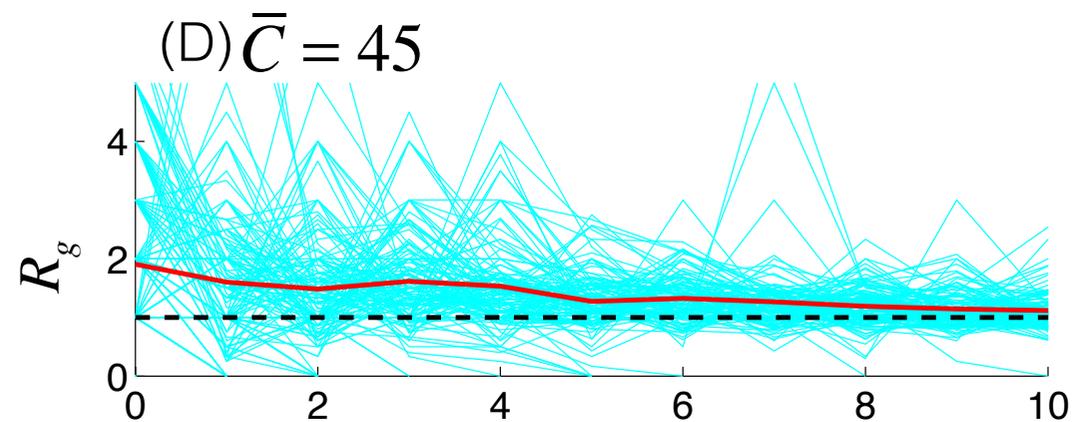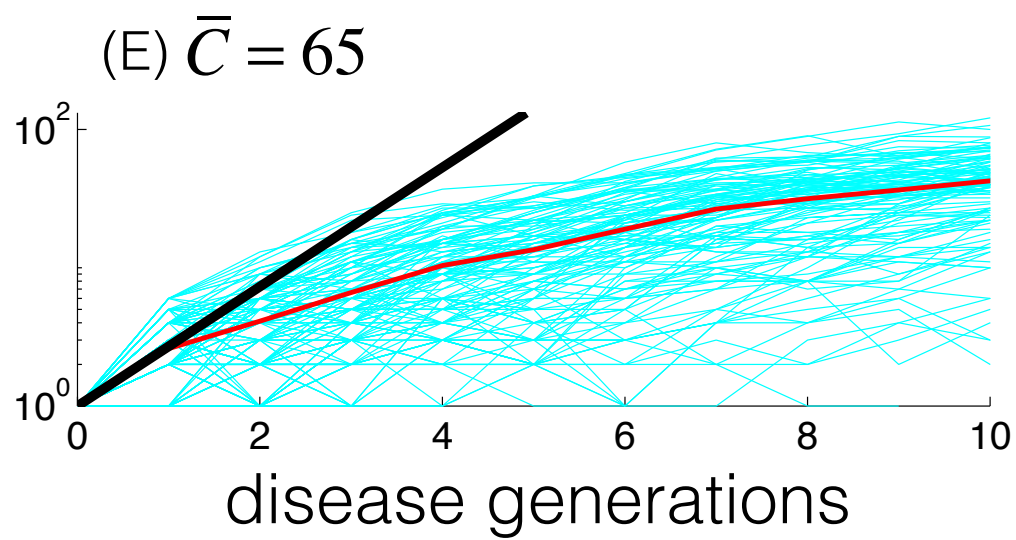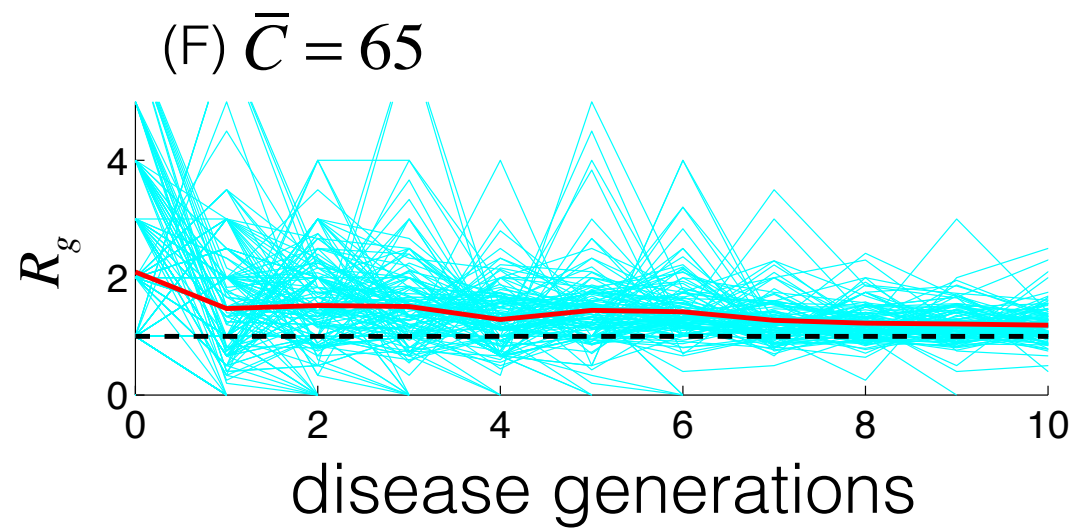

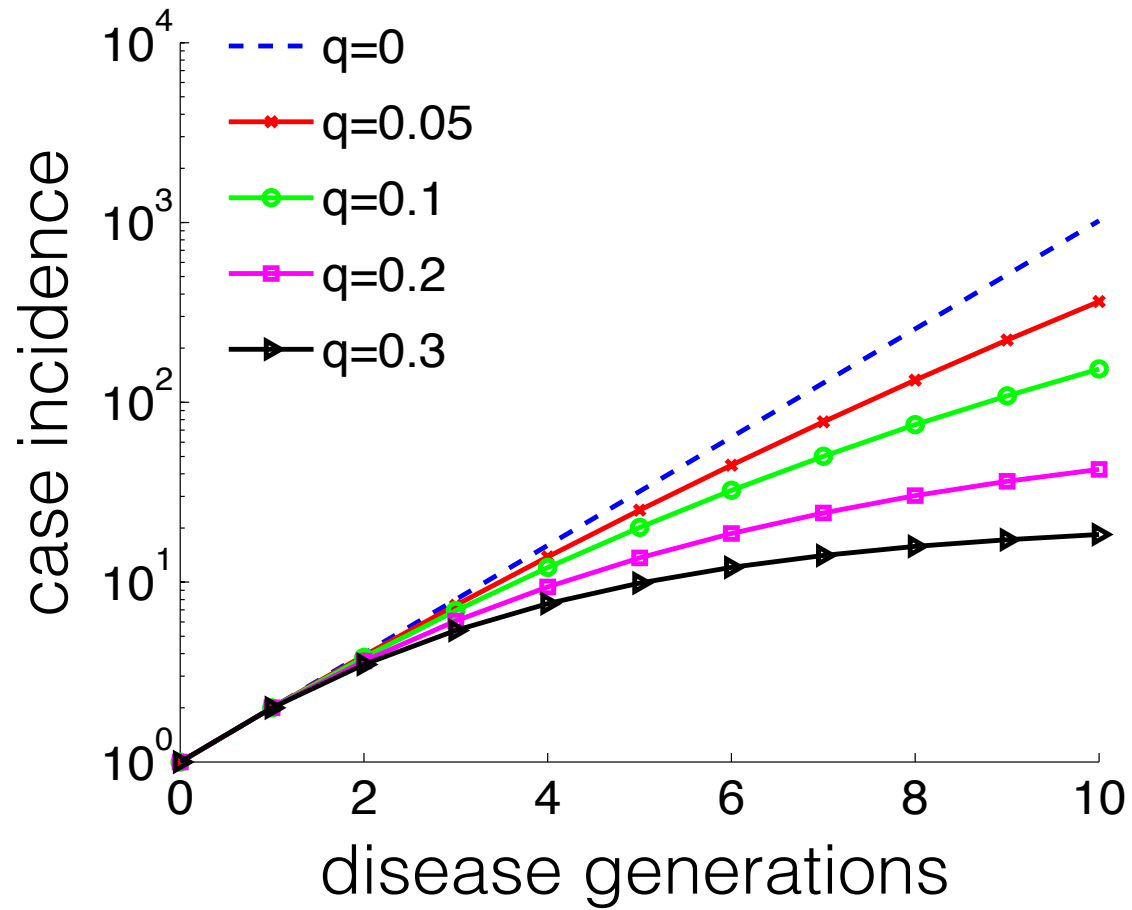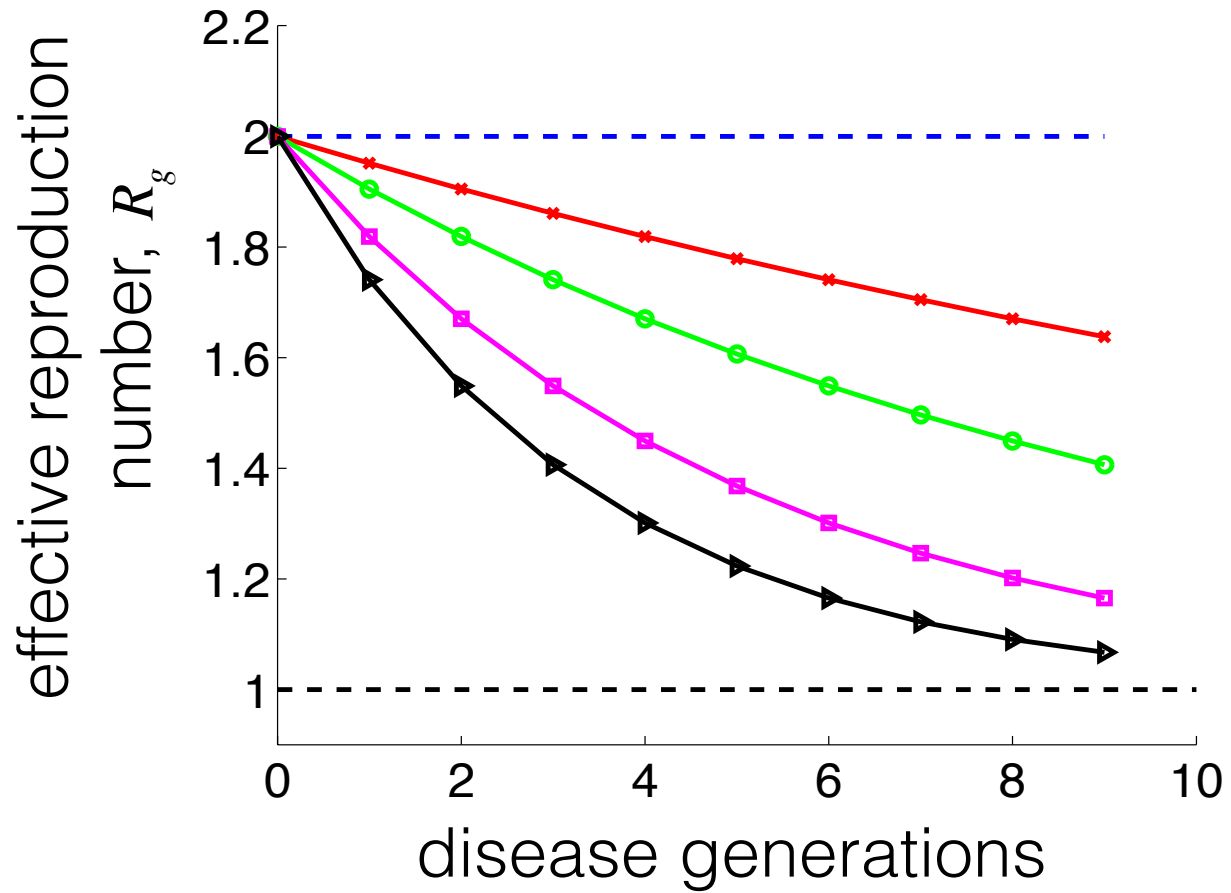

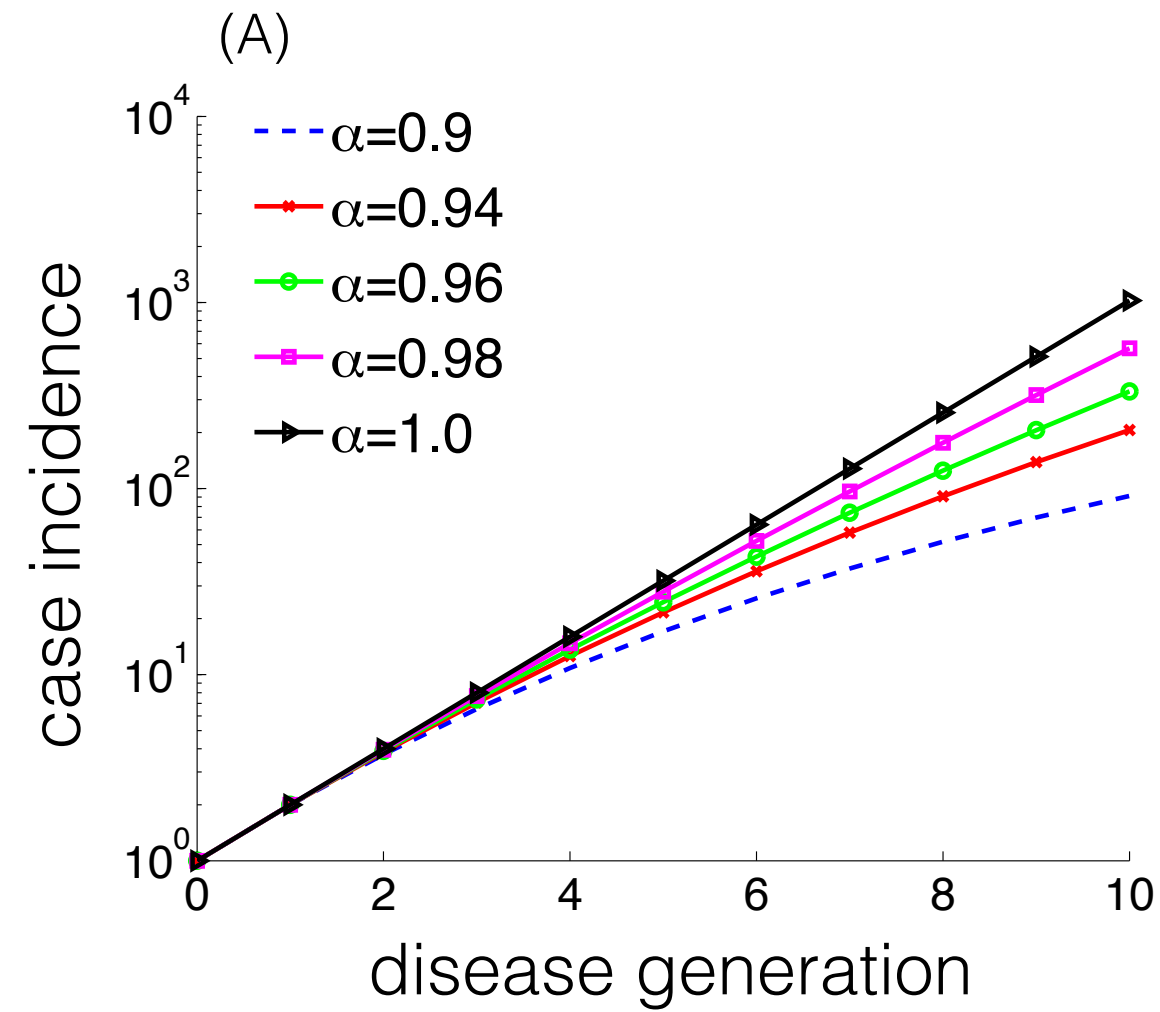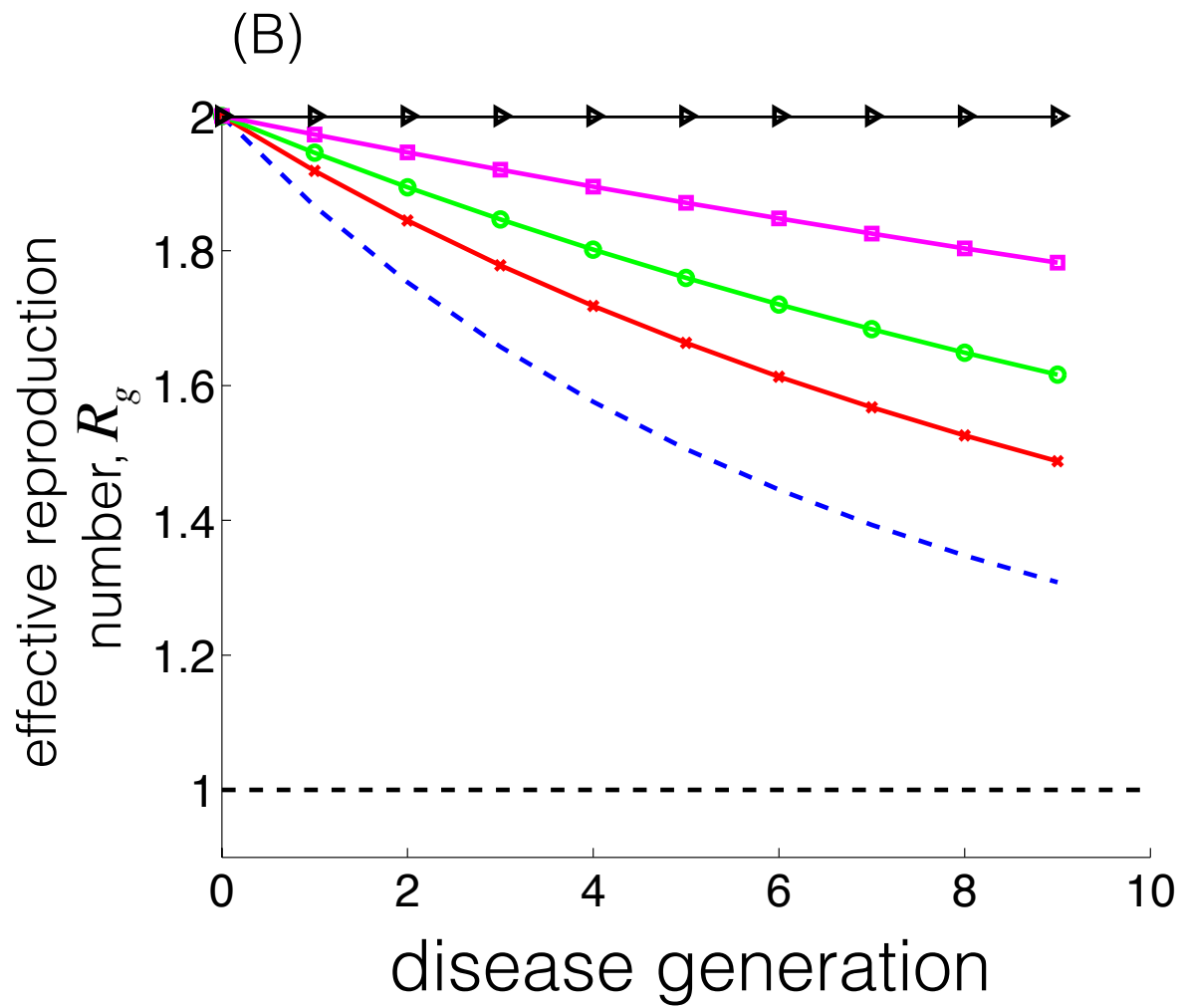

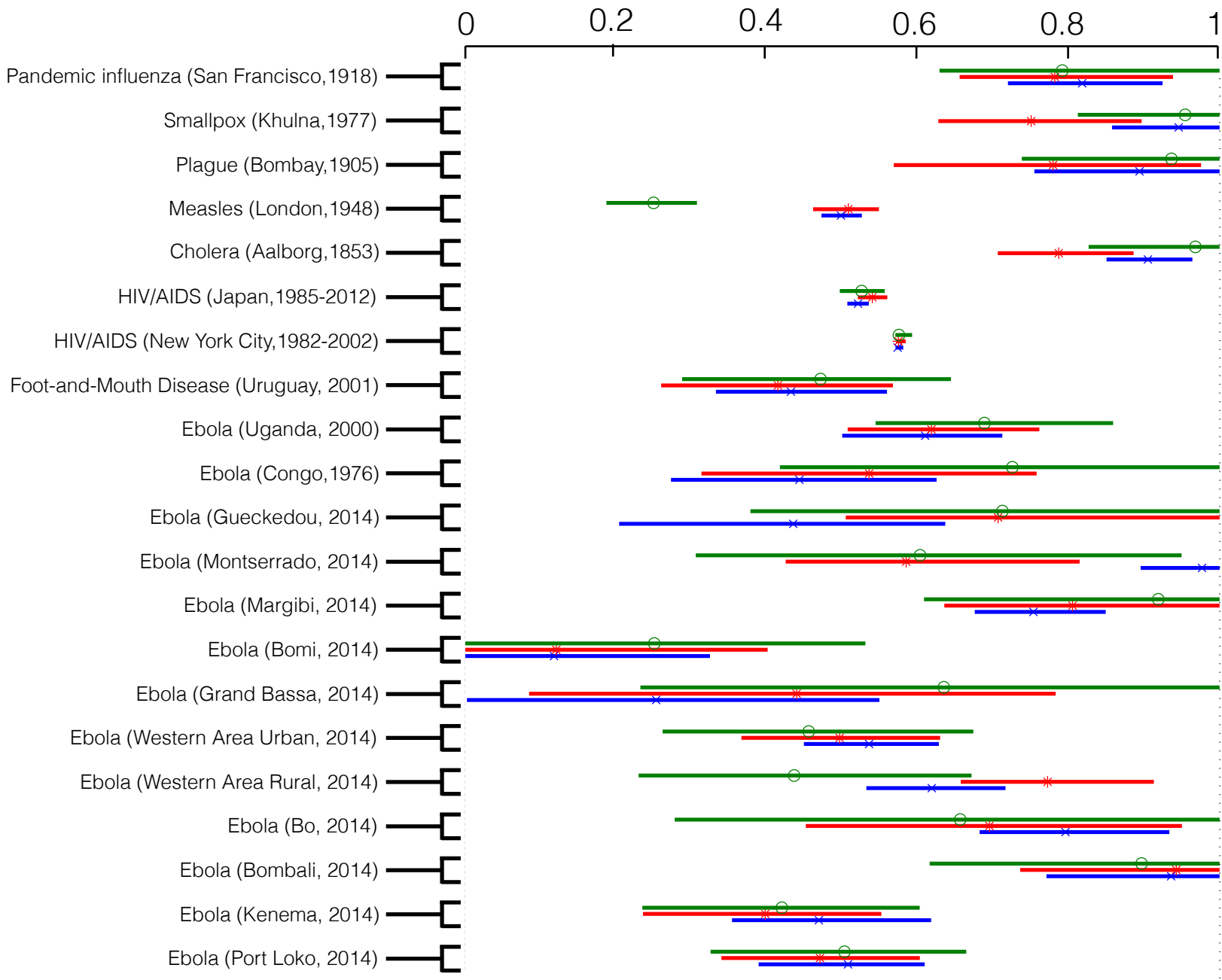

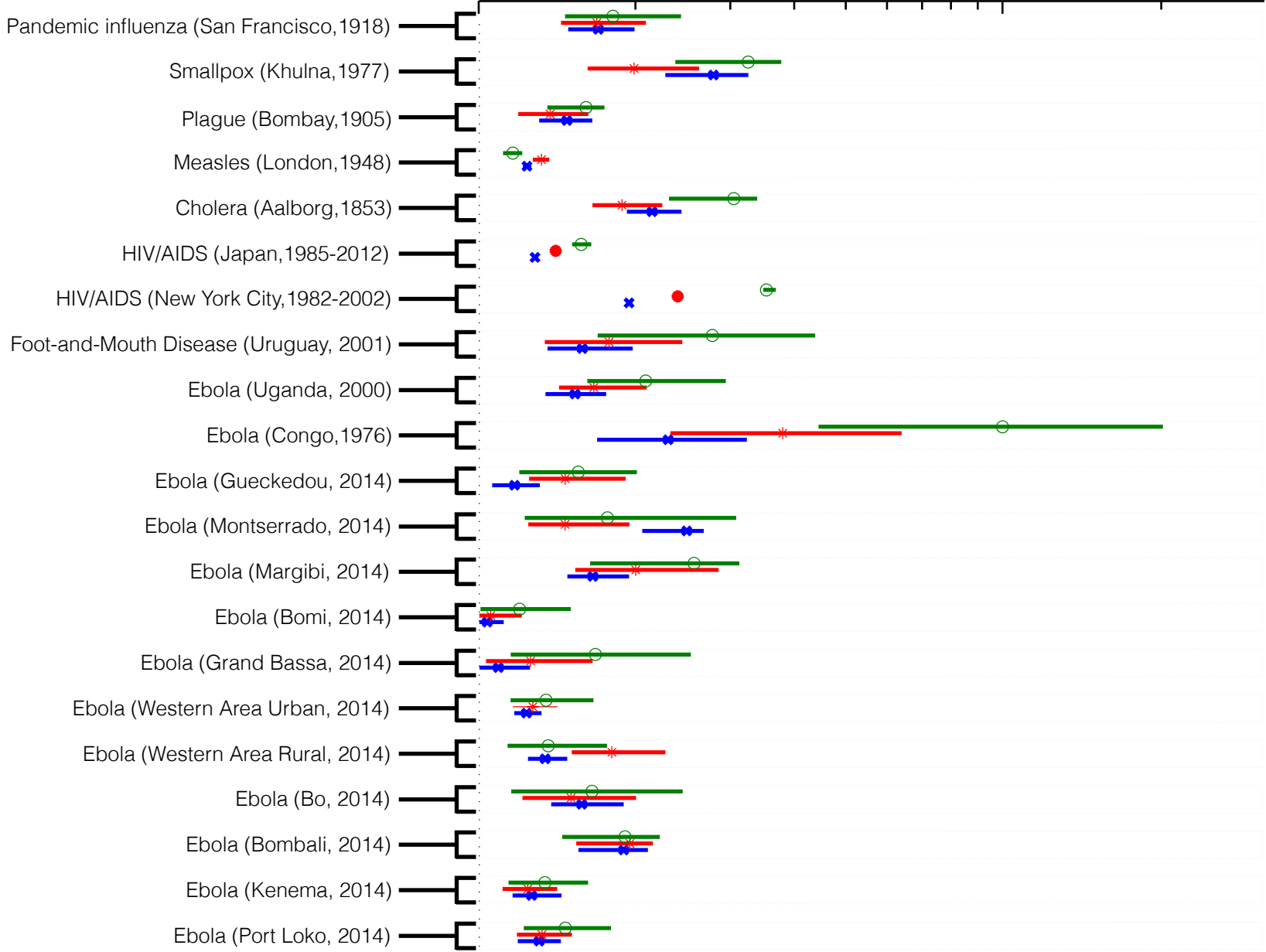

# Supporting Information

**Characterizing the reproduction number of epidemics with early sub-exponential growth dynamics**[1]

Gerardo Chowell[1,2], Cécile Viboud[2], Lone Simonsen[3,4], Seyed M. Moghadas[5]

[1] School of Public Health, Georgia State University, Atlanta, GA, USA

[2] Division of International Epidemiology and Population Studies, Fogarty International Center, National Institutes of Health, Bethesda, MD, USA

[3] Department of Public health, University of Copenhagen, Copenhagen, Denmark

[4] Department of Global Health, School of Public Health and Health Services, George Washington University, Washington DC, USA

[5] Agent Based Modelling Laboratory, York University, Toronto, Canada.

---

*Alternative formulation of the generalized-growth model (GGM)*

Here we derive an alternative formulation of the generalized-growth model where the biological factor acts on the growth rate $r$ rather than $C$. The original generalized-growth model is given by [1]:

$$C' = rC^p \qquad (2.1)$$

where $C'(t)$ describes the incidence curve over time $t$, the solution $C(t)$ describes the cumulative number of cases at time $t$, $r$ is a positive parameter denoting the growth rate (with the units of (people)$^{1-p}$ per time), and $p \in [0,1]$ is a 'deceleration of growth' parameter (dimensionless).

We first apply a change of variable $u = \log(C)$ so that $C = e^u$. Then the model becomes:

$$u' = re^{u(p-1)}$$

The solution of this model is given by:

$$u(t) = \begin{cases} rt & p = 1 \\ \dfrac{1}{1-p}\log\!\left(r(1-p)t + e^{u_0(1-p)}\right) & 0 \leq p < 1 \end{cases}$$

where $u_0 = \log(C_0)$. The rate of change of the growth rate in the original generalized-growth model (Equation 1) is given by:

$$u'(t) = \begin{cases} r & p = 1 \\ \dfrac{r}{r(1-p)t + e^{u_0(1-p)}} & 0 \leq p < 1 \end{cases}$$

Hence, the alternative formulation of the generalized-growth model is given by:

$$C'(t) = u'(t)C(t)$$

where $u'(t) = \begin{cases} \dfrac{r}{r(1-p)t + e^{\log(C_0)(1-p)}} & 0 \leq p < 1 \\ r & p = 1 \end{cases}$

The same numerical results are obtained from both formulations of the generalized growth model as shown in Figure S1.

**Figure S1**. Comparison of the numerical results derived from both formulations of the generalized-growth model for different values of $p$ when $r = 1.5$ and $C_0 = 1$.

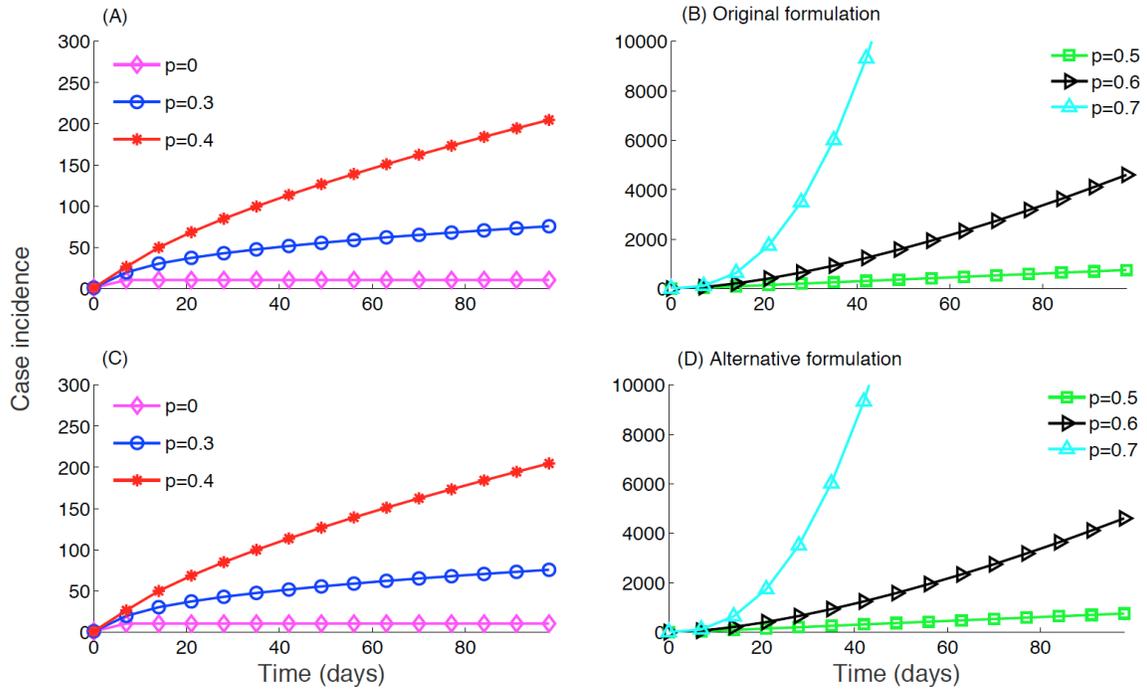

*Mathematical proof for* $R_g^{subexp} \to e^{rT_g}$ *as* $p \to 1^-$

Using methods of limit, one can easily show that the reproduction number $R_g^{subexp}$ according to disease generations converges to $e^{rT_g}$ as $p \to 1^-$ as follows:

$$R_g^{subexp} = \left[1 + \frac{r(1-p)T_g}{r(1-p)gT_g + C_0^{1-p}}\right]^{\frac{p}{1-p}} \qquad (2.6)$$

If $p \to 1^-$ then $R_g^{subexp} \approx 1^\infty$

Taking the logarithm on both sides we obtain (when $C_0 \neq 0$):

$$\ln(R_g^{subexp}) = \frac{p}{1-p} \ln\left[1 + \frac{r(1-p)T_g}{r(1-p)gT_g + C_0^{1-p}}\right] \approx \frac{0}{0} \text{ when } p \to 1^-$$

Applying L'Hospital's Rule, we obtain:

$$y = \frac{1\left[\frac{-rT_g[1] - \frac{0}{1^2}}{1+0}\right]^0}{-1} = \frac{-rT_g}{-1} = rT_g$$

as $p \to 1^-$. Then $\lim_{p \to 1^-} R_g^{subexp} = e^{rT_g}$

Putting everything together, we have:

$$R_g^{subexp} = \begin{cases} 1 & \text{If } g \to \infty \text{ and } p < 1 \\ e^{rT_g} & \text{If } p \to 1^- \end{cases}$$

**Figure S2**. Simulated profiles of the effective reproduction number during the first 5 generation intervals derived from case incidence curves of the generalized-growth model with different values of the growth rate (r) and the deceleration of growth parameter (p). The initial number of cases is set to C(0)=1. Estimates of the effective reproduction number are generated assuming a fixed generation interval at 3 days.

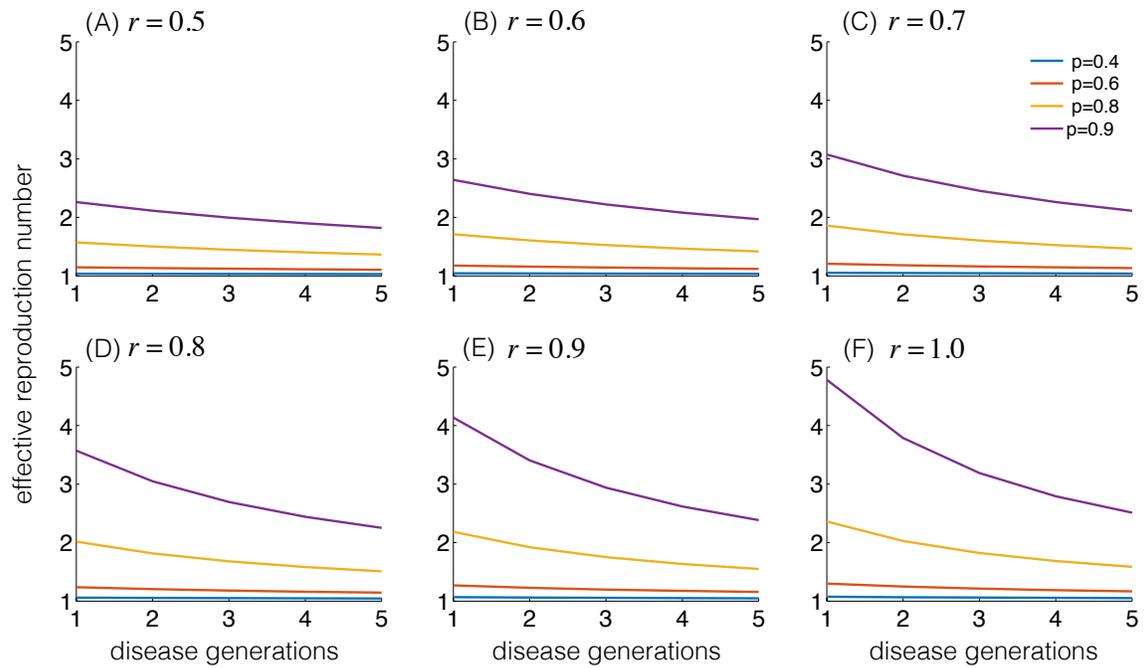

**Figure S3.** Simulations of the early epidemic growth phase derived from the SIR model described by Equations (2.10) for different values of the power-law scaling parameter $\alpha$, $\gamma = \frac{1}{5}$, and (A) $\beta_0 = 0.4$, (B) $\beta_0 = 0.48$, (C) $\beta_0 = 0.56$, and (D) $\beta_0 = 0.6$ with a large population size N set at $10^8$. The epidemic simulations start with one infectious individual. In semi-logarithmic scale, exponential growth is evident if a straight line fits well several consecutive disease generations of the epidemic curve, whereas a strong downward curvature in semi-logarithmic scale is indicative of sub-exponential growth. Our simulations show that case incidence curves display early sub-exponential growth dynamics even for values of $\alpha$ slightly below 1.0.

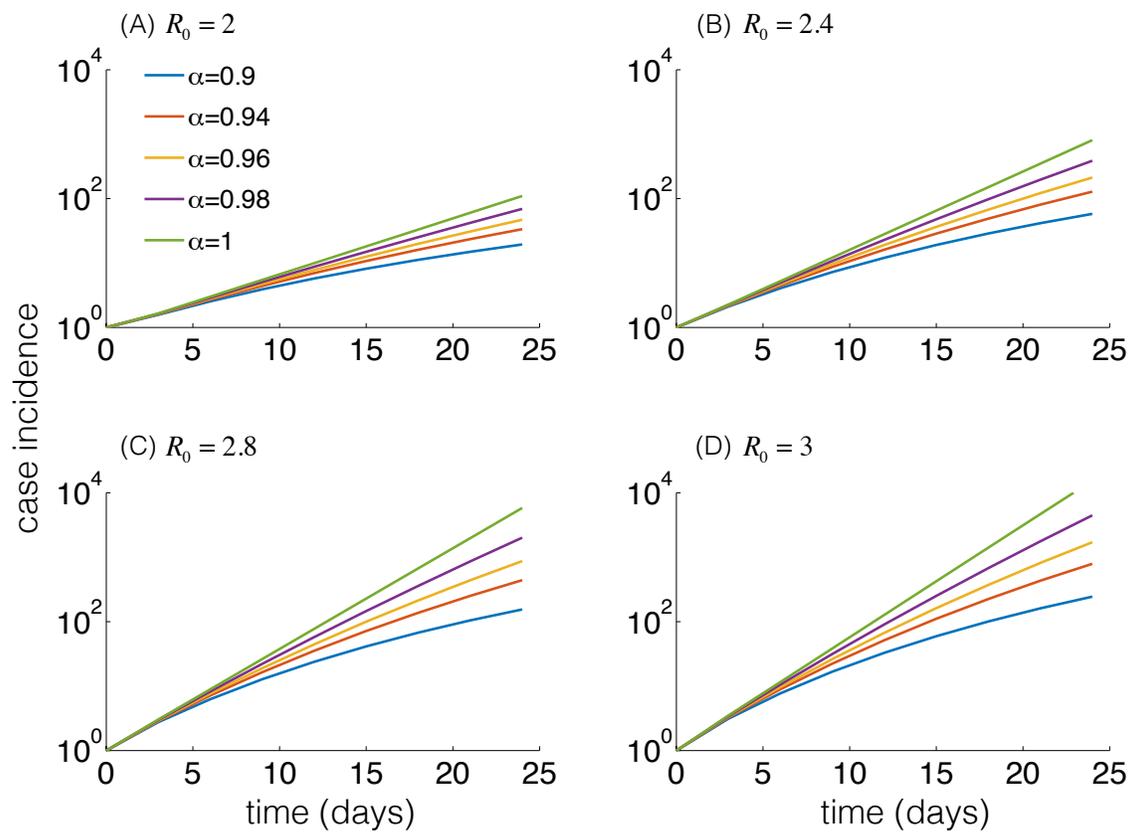

**Figure S4.** Examples of the fits provided by the exponential and generalized-growth models to three infectious disease outbreaks (pandemic influenza in San Francisco in 1918, HIV/AIDS epidemic in Japan, and Ebola epidemic in Western Urban, Sierra Leone in 2014).

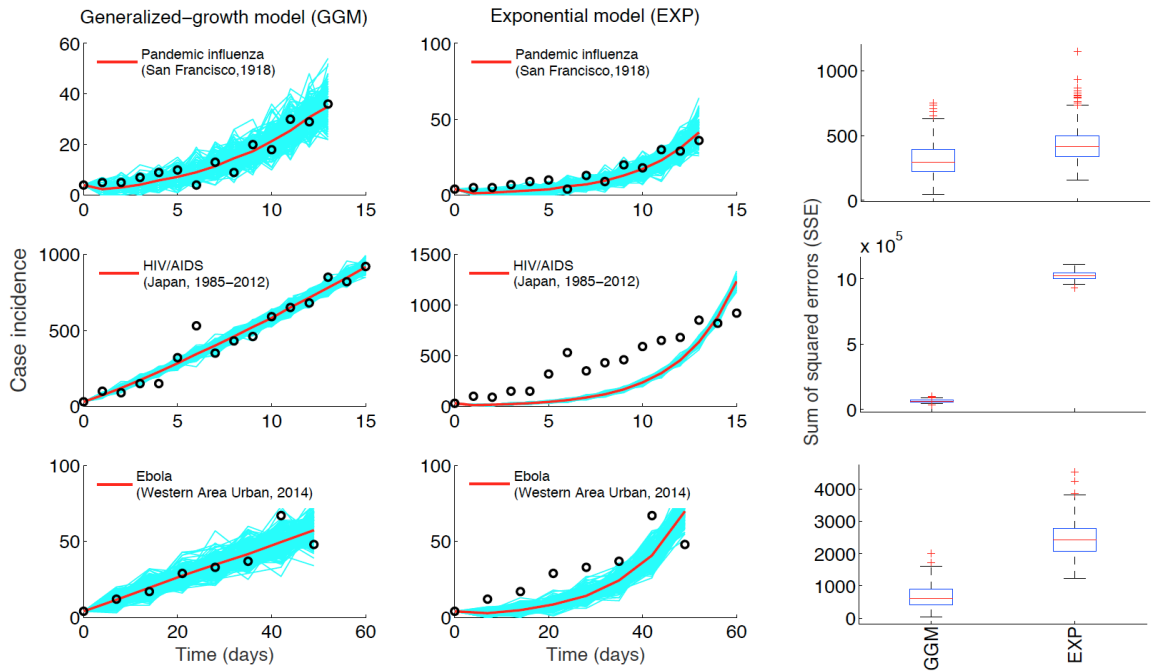

**Figure S5**. Comparison of the goodness of fit provided by the exponential and the generalized-growth models across all of the 21 infectious disease outbreaks (see Table 1).

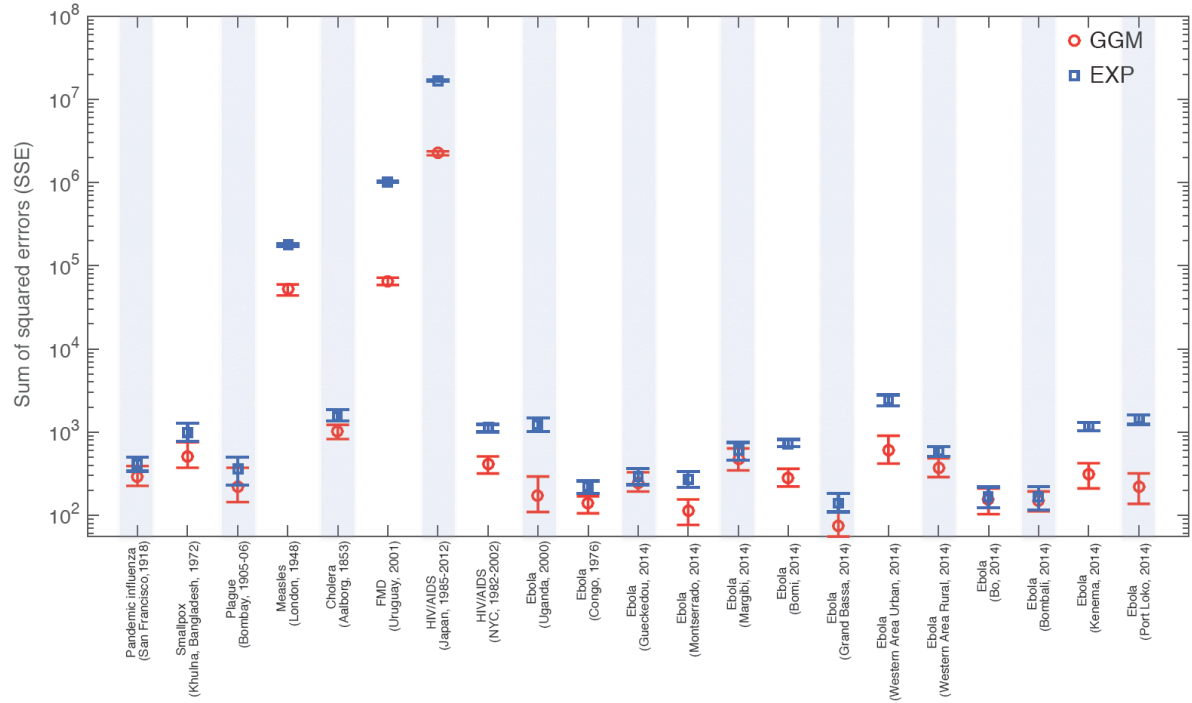

**Figure S6.** Mean estimates of the effective reproduction number and the deceleration of growth parameter $p$ derived from our sample of infectious disease datasets (Table 1) were significantly correlated for three estimation periods with an initial phase length comprising 3 to 5 generation intervals.

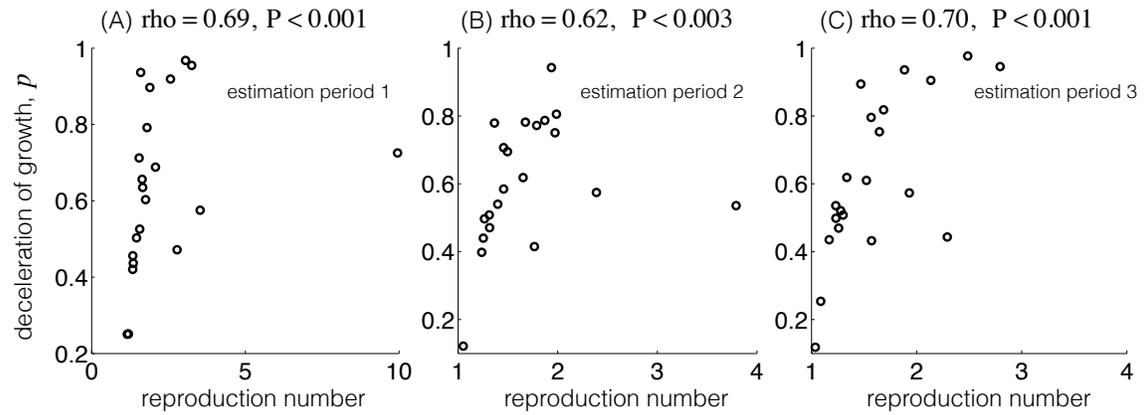

**Figure S7. The 1918 influenza pandemic in San Francisco.** Estimates and 95% confidence intervals of the effective reproduction number derived from fitting the generalized-growth model to an increasing length of the early epidemic phase comprising of approximately 3-5 disease generation intervals. The generation interval is assumed to be gamma distributed with the mean of 3 days and standard deviation of 1 day. Estimates and 95% confidence intervals for parameters $r$ and $p$ are also shown.

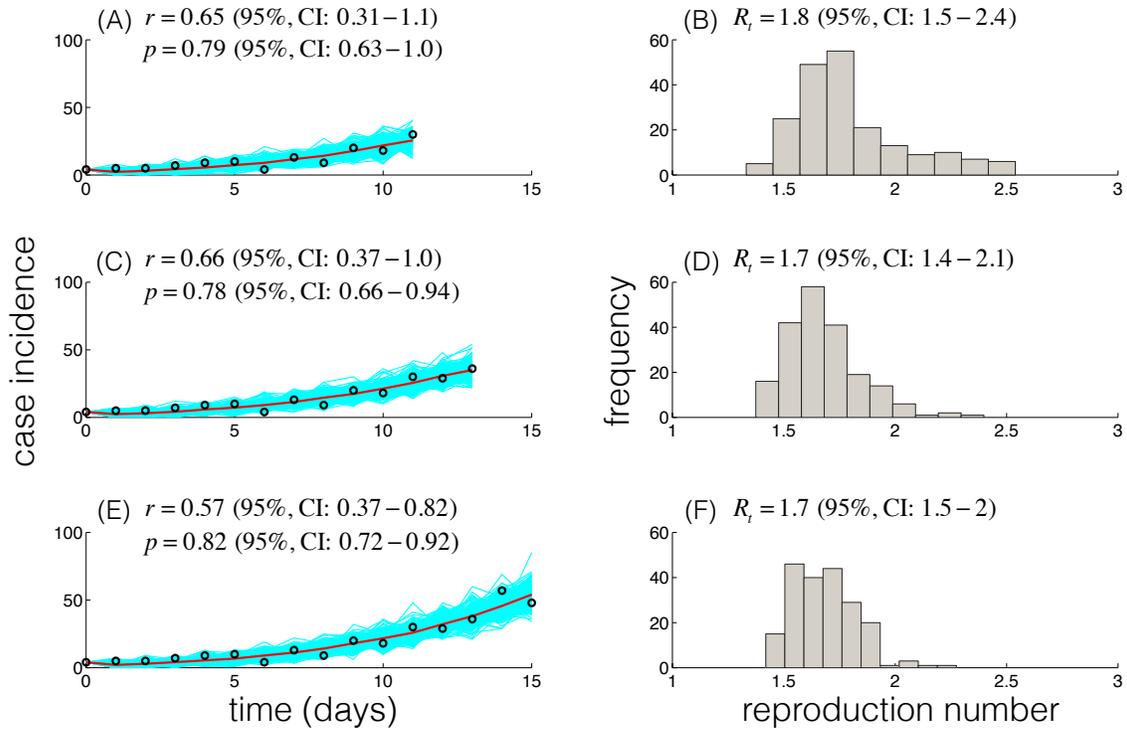

**Figure S8. The 2001 foot-and-mouth disease epidemic in Uruguay.**
Estimates and 95% confidence intervals of the effective reproduction number derived from fitting the generalized-growth model to an increasing length of the early epidemic phase comprising of approximately 3-5 disease generation intervals. The generation interval is assumed to be gamma distributed with the mean of 5 days and standard deviation of 1 day. Estimates and 95% confidence intervals for parameters $r$ and $p$ are also shown.

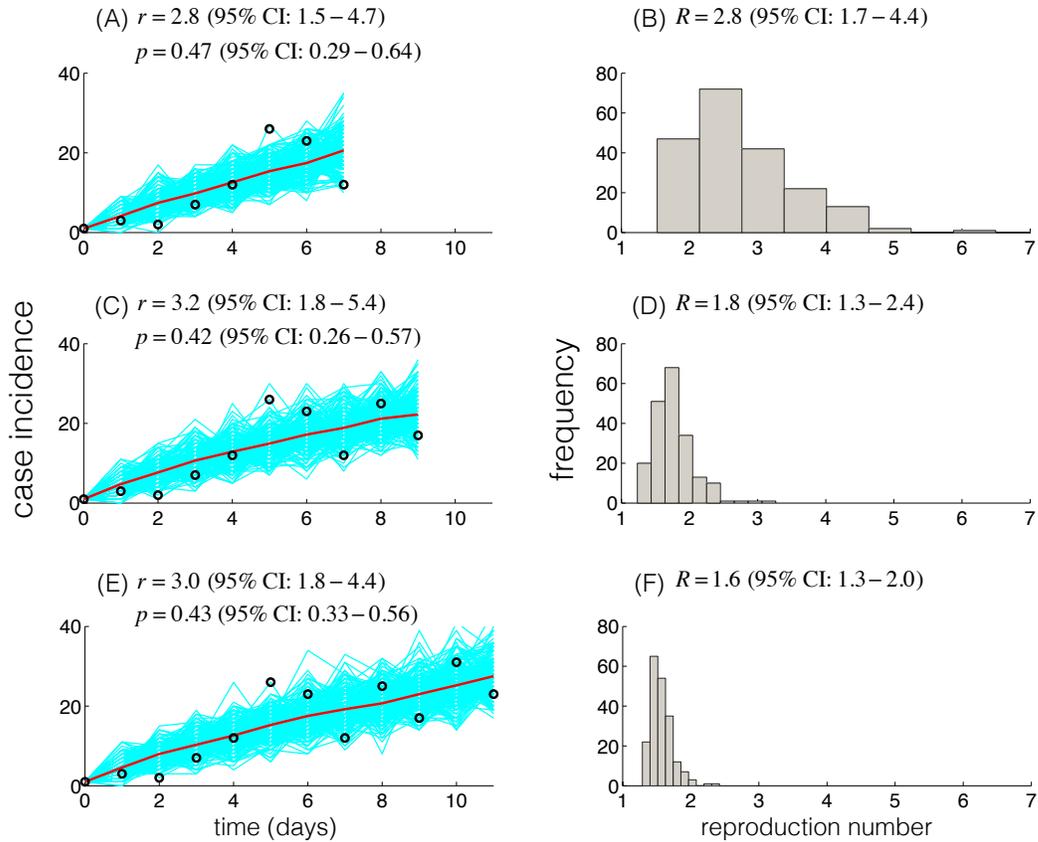

**Figure S9. The HIV/AIDS epidemic in Japan (1985-2012).**
Estimates and 95% confidence intervals of the effective reproduction number derived from fitting the generalized-growth model to an increasing length of the early epidemic phase comprising of approximately 3-5 disease generation intervals. The generation interval is assumed to be gamma distributed with the mean of 4 years and standard deviation of 1.4 years. Estimates and 95% confidence intervals for parameters $r$ and $p$ are also shown.

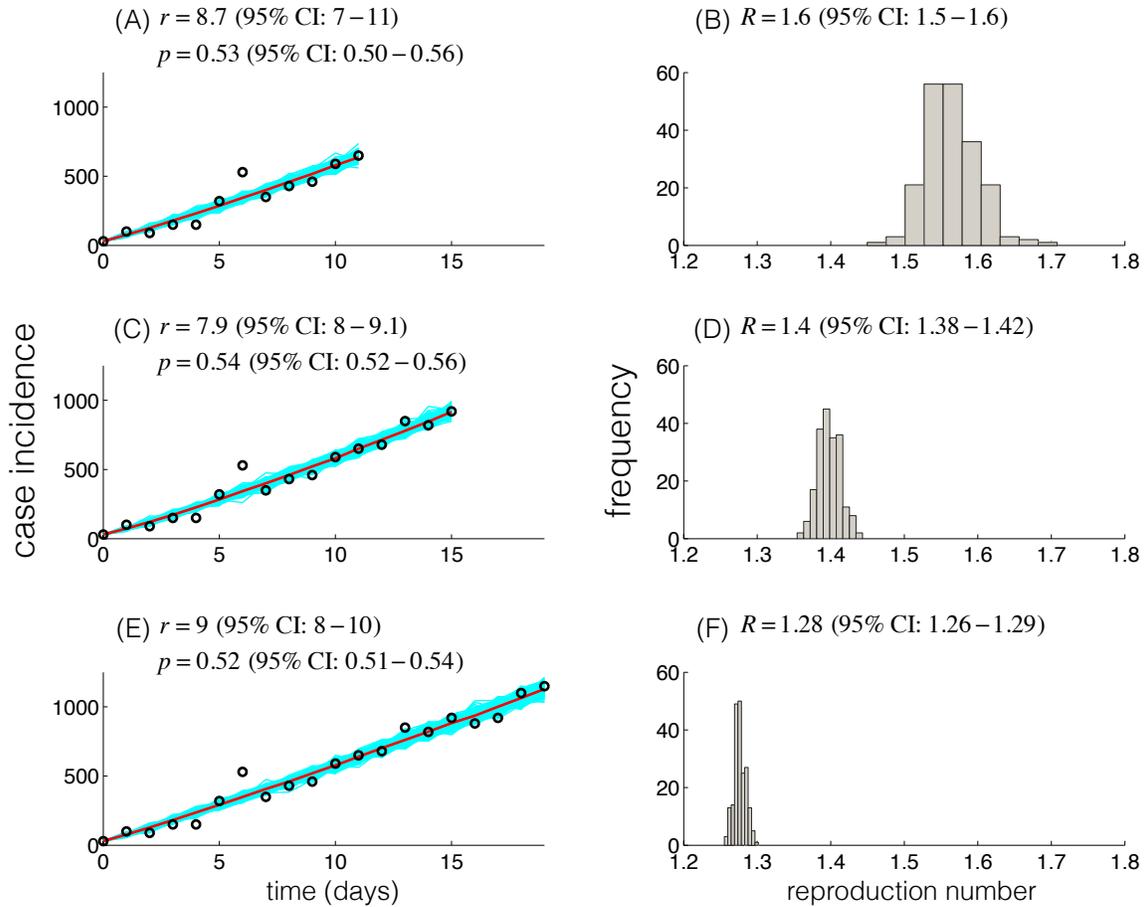

**Figure S10. The 2014-15 Ebola epidemic in Gueckedou, Guinea.**
Estimates and 95% confidence intervals of the effective reproduction number derived from fitting the generalized-growth model to an increasing length of the early epidemic phase comprising of approximately 3-5 disease generation intervals. The generation interval is assumed to be gamma distributed with the mean of 19 days and standard deviation of 11 days. Estimates and 95% confidence intervals for parameters $r$ and $p$ are also shown.

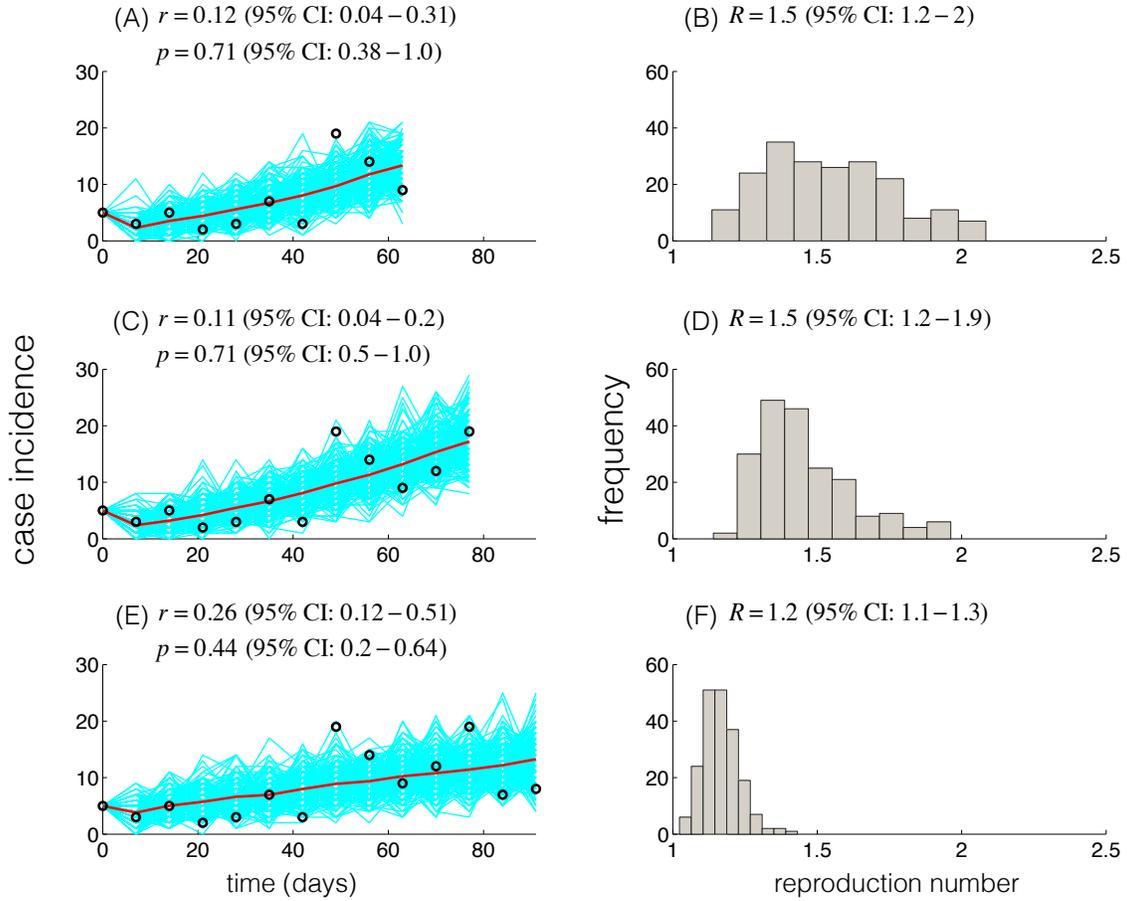

**Figure S11. The 2014-15 Ebola epidemic in Western Area Urban, Sierra Leone.** Estimates and 95% confidence intervals of the effective reproduction number derived from fitting the generalized-growth model to an increasing length of the early epidemic phase comprising of approximately 3-5 disease generation intervals. The generation interval is assumed to be gamma distributed with the mean of 11.6 days and standard deviation of 5.6 days. Estimates and 95% confidence intervals for parameters $r$ and $p$ are also shown.

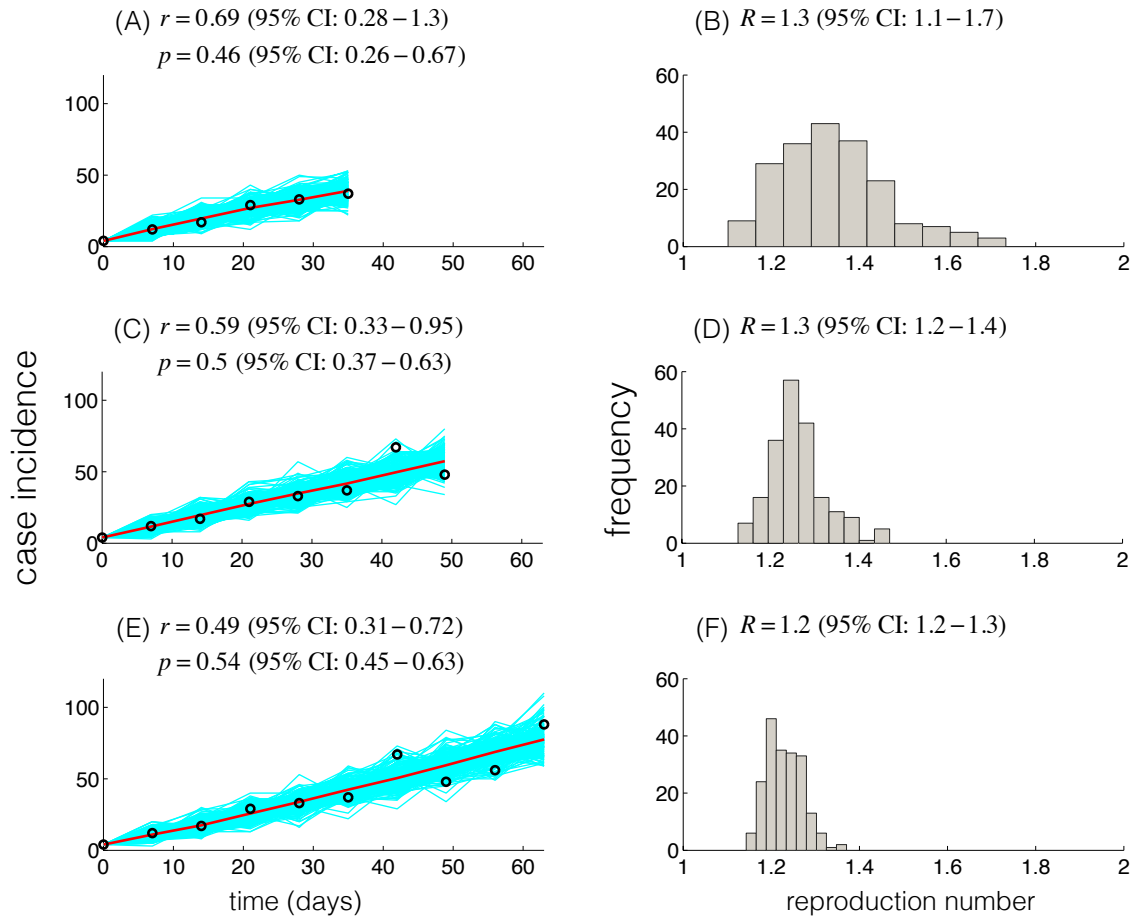